\documentclass[journal=jpcafh,manuscript=article]{achemso}
\usepackage{longtable}
\usepackage{multirow}
\usepackage{amsmath}
\usepackage{subfigure}
\usepackage{soul}
\usepackage{bm}
\usepackage[normalem]{ulem}

\usepackage{amssymb}
\usepackage{rotating}
\makeatletter
\newcommand*{\addFileDependency}[1]{
  \typeout{(#1)}
  \@addtofilelist{#1}
  \IfFileExists{#1}{}{\typeout{No file #1.}}
}
\makeatother

\usepackage{xcolor}

\usepackage{multicol} \usepackage{wrapfig} \usepackage{enumitem}
\usepackage{bm} \usepackage{outlines} 
\SectionNumbersOn

\author{Sugata Goswami} \affiliation{Department of Chemistry,
  University of Basel, Klingelbergstrasse 80, CH-4056 Basel,
  Switzerland} \altaffiliation{Present address: Department of Chemistry, Medi-Caps
  University, A.B. Road, Pigdamber, Indore - 453 331 (M.P.), India}

\author{Silvan K\"aser} \affiliation{Department of Chemistry,
  University of Basel, Klingelbergstrasse 80, CH-4056 Basel,
  Switzerland}

\author{Raymond J. Bemish} \affiliation{Air Force Research Laboratory,
  Space Vehicles Directorate, Kirtland AFB, New Mexico 87117, USA}

\author{Markus Meuwly} \email{m.meuwly@unibas.ch}
\affiliation[University of Basel] {Department of Chemistry, University
  of Basel,\\ Klingelbergstrasse 80, 4056 Basel, Switzerland}

\title{On the Effect of Aleatoric and Epistemic Errors on the
  Learnability and Quality of NN-based Potential Energy Surfaces}

\begin{document}

\date{\today}\\

\begin{abstract}
The effect of noise in the input data for learning potential energy
surfaces (PESs) based on neural networks for chemical applications is
assessed. Noise in energies and forces can result from aleatoric and
epistemic errors in the quantum chemical reference calculations.
Statistical (aleatoric) noise arises for example due to the need to
set convergence thresholds in the self consistent field (SCF)
iterations whereas systematic (epistemic) noise is due to, {\it inter
  alia}, particular choices of basis sets in the calculations. The two
molecules considered here as proxies are H$_{2}$CO and HONO which are
examples for single- and multi-reference problems, respectively, for
geometries around the minimum energy structure. For H$_2$CO it is
found that adding noise to energies with magnitudes representative of
single-point calculations does not deteriorate the quality of the
final PESs whereas increasing the noise level commensurate with
electronic structure calculations for more complicated,
e.g. metal-containing, systems is expected to have a more notable
effect. However, the effect of noise on the forces is more
noticeable. On the other hand, for HONO which requires a
multi-reference treatment, a clear correlation between model quality
and the degree of multi-reference character as measured by the $T_1$
amplitude is found. It is concluded that for chemically "simple" cases
the effect of aleatoric and epistemic noise is manageable without
evident deterioration of the trained model - although the quality of
the forces is important. However, considerably more care needs to be
exercised for situations in which multi-reference effects are present.
\end{abstract}

\section{Introduction}
Machine learning (ML) has been established as a promising tool for
representing inter- and intramolecular potential energy surfaces (PES)
for applications in chemistry, physics, and
biophysics.\cite{unke:2021,manzhos:2020,MM:2021} The great interest in
developing and using such statistical models\cite{vapnik:1998} stems
from the fact that - computationally expensive - energies (and forces)
from electronic structure calculations can now be obtained at high
levels of theory for large numbers of geometries whereas evaluation
times for energies and forces of trained statistical models are
comparatively short. As an example, for tri- and tetra-atomic
molecules energies for $\sim 10^4$ geometries at the multi reference
configuration interaction (MRCI) or coupled cluster singles doubles
and perturbative triples (CCSD(T)) levels of theory can be determined
routinely using at least aug-cc-pVTZ basis set
quality.\cite{MM.cno:2018,MM.co2:2021} For larger molecules,
containing up to 10 heavy atoms, energies based on MP2 or density
functional theory (DFT) calculations using slightly smaller basis sets
for $\sim 10^5$ geometries are routinely
affordable.\cite{K_ser_2020,Chenpccp2021}\\

\noindent
An important aspect in the numerical treatment of chemical systems is
the role played by inaccuracies inherent to such an approach.  Every
computational treatment has only finite accuracy and managing this
aspect is important. The role of "noise" in machine-learned models has
been considered from a number of perspectives. For one, it has been
proposed to adapt the cost function to be minimized to account for
errors in the input data.\cite{pintelon:2000} The cost function
employed determines the noise-sensitivity of the model and is
therefore a meaningful way to account for uncertainty on the input
data. An alternative is to explicitly model the noise as an additional
layer in the NN-architecture.\cite{goldberger:2016} Related to this,
noise was included directly into the training ("training-with-noise
algorithm"). Within this framework it was found that injecting
measured amounts of noise of a particular structure can even improve
the generalization capabilities for classification
tasks.\cite{gardner:1989,benedetti:2023} Finally, from the perspective
of the type of uncertainty, aleatoric (statistical) and epistemic
(systematic) uncertainties for the output of a trained NN have been
analyzed for function estimation and
classification.\cite{valdenegro:2022}.\\

\noindent
In the present work the role of noise in the input data is studied and
analyzed for training multi-dimensional PESs based on reference {\it
  ab initio} calculations. Electronic structure calculations in
general require certain input from the user that controls the level of
convergence of a calculation for a given conformation. For one, the
self consistent field (SCF) calculation needs to be converged
according to one or several properties, including the root mean
squared (RMS) or the maximum change in the density matrix. This limits
the aleatoric uncertainty of a calculation but does not influence the
epistemic uncertainty. Hence, depending on the thresholds chosen at
this step the final energy and associated forces for a given structure
differ. This variability in the total energy induced, e.g., by
particular choices of user-defined parameters is referred to as noise
in the present work.\\

\noindent
Depending on how complicated the electronic structure of a molecule
is, more or less strict convergence criteria can be applied to the
wavefunction at the SCF level. For closed-shell, single-reference
species the tightest criteria possible are usually used whereas
molecules containing transition metals can pose severe problems in
terms of converging the reference wavefunction for a given
geometry. Additional measures, such as ``level
shifting''\cite{saunders1973level} may be required to converge
energies for individual or even all conformations of interest. Again,
this limits aleatoric but not epistemic uncertainty. Hence,
determining the reference dataset for a full-dimensional, reactive PES
for a species with complicated electronic structure can become a
daunting task whereby individual geometries require dedicated
adjustments of system parameters. This leads to variable aleatoric
uncertainty across the reference data which is propagated into the
model building and further influences the epistemic uncertainty of the
final statistical model. Epistemic uncertainty arises due to the need
to choose a particular basis set or to select a specific quantum
chemical method which is an approximation to a full configuration
interaction (FCI) treatment.\\

\noindent
Such aspects become even more relevant when PESs are used in broader
dynamics studies of chemical reactions. Using noisy reference data to
represent the PES either as a neural network (NN) or with kernel-based
methods, can lead to difficulties in training the statistical
model. It has, for example, been found that using reproducing kernel
Hilbert space (RKHS) representations on electronic structure
calculations for He--H$_2^+$ at the full CI level work very reliably
whereas for the same grid with energies determined from MRCI+Q
calculations regularization is needed for a smooth
representation.\cite{MM.heh2:2019} Hence, depending on the underlying
quantum chemical method used the learnability differs and the
resulting model is of higher or
inferior quality. Such uncertainty will propagate through the entire
pipeline which, in that case, consisted of running dynamics
simulations on the PES and determining
observables.\cite{MM.no2:2020,MM.n2o:2020,MM.co2:2021}\\

\noindent
 It is the aim of the present work to investigate and to quantify the
 effect of aleatoric and epistemic uncertainties in reference data on
 training machine-learned PESs. This is (partially) motivated by
 earlier findings in which learning on CCSD(T)-F12 data (energies and
 forces) led to a rapid convergence toward an error floor of several
 $10^{-4}$ kcal/mol while a lower out-of-sample error was achieved
 using \textit{ab initio} data of MP2 quality.\cite{MM.h2co:2020}
 Inspection of the literature shows that the gradients in MOLPRO at
 the CCSD(T)-F12 level indeed can be less accurate than machine
 precision.\cite{gyHorffy2018analytical} Additionally, the work
 explores the effects of convergence criteria required to define in
 quantum chemical calculations and of using single-reference methods
 for systems with multi-reference character on the \textit{ab initio}
 data and subsequent learning. The two molecules considered here are
 H$_2$CO and HONO.\cite{kasting:1993,finlayson:1986} Both molecules
 are relevant in atmospheric chemistry and serve as benchmark systems
 for single- (H$_2$CO) and multi- (HONO) reference systems from an
 electronic structure perspective.\\

\noindent
The work is organized as follows: first, the methods employed are
described. This is followed by results on learning with perturbed
energies and with perturbed forces. The role of hyperparameters is
briefly considered followed by an exploration of perturbations due to
convergence of the electronic structure calculations and
multi-reference effects.

\section{Methods}
This section describes the generation of the datasets and the learning 
protocols used in the present work. The two molecules considered
are H$_2$CO and HONO. For H$_2$CO a RKHS representation of reference
MP2/aug-cc-pVTZ data is available\cite{MM.h2co:2020} from which
``clean'' energies and forces - apart from numerical imprecisions -
can be determined. For HONO a recent study\cite{MM.tl:2021} suggested
that training PhysNet on MP2 data - neglecting multi-reference effects
in the electronic structure calculations - makes training the NN more
cumbersome. Whether or not this observation is related to the neglect of
multi-reference effects will be quantitatively assessed in the present
work.\\

\subsection{Clean Models and Models with Noise for H$_2$CO}
For H$_2$CO clean datasets were obtained by evaluating a RKHS
representation for 3601 reference geometries.\cite{MM.tl:2021} Here,
clean refers to the notion that apart from numerical imprecisions no
further sources of noise arise because for a given geometry the energy
is obtained from evaluation of a function based on a matrix-vector
multiplication. The kernel coefficients of the
RKHS-based\cite{ho1996general,MM.rkhs:2017} PES, available from
previous work,\cite{MM.h2co:2020} were generated by using energies and
forces calculated at the MP2/aug-cc-pVTZ level of theory.\\

\noindent
To evaluate aleatoric uncertainty arising from effects such as
convergence thresholds and/or multi-reference effects in electronic
structure calculations, Gaussian-distributed noise with given
amplitude was generated and added according to the following
procedure. Gaussian random numbers were generated from a distribution
with zero mean and three standard deviation (SD = $10^{-5}$, $10^{-6}$
and $10^{-7}$ eV (for energy) and eV/\AA\/ (for forces)). For each
data point of the clean PES, energy and force perturbations were drawn
from the distributions and added to the data point. This leads to six
sets of perturbed energies and forces which are used in addition to
the clean dataset to learn representations of the PES using
PhysNet.\cite{MM.physnet:2019}\\

\subsection{Machine-Learned Potential Energy Surfaces}
For all datasets generated and employed in the present work, PhysNet
was used to learn a representation of the PES.\cite{MM.physnet:2019}
PhysNet is a high-dimensional, message passing NN built to learn
molecular properties such as energy and forces from \textit{ab initio}
data.\cite{MM.physnet:2019} Starting from the Cartesian coordinates
$\bm{r}_{i}$ and nuclear charges $z_{i}$ of all atoms $i$ of an
arbitrary molecular geometry a feature vector is learned to best
describe the atoms' local chemical environment. Atomic energy
contributions are then predicted based on the feature vector. The
learnable parameters of the NN-based PES are determined by minimizing
a suitable loss function
\begin{equation}
 \mathcal{L} = w_{\rm E}|E-E^{\rm ref}| + \frac{w_{\rm
     F}}{3N}\sum_{i=1}^{N}\sum_{\alpha=1}^{3}\Big|-\frac{\partial
   E}{\partial r_{i,\alpha}}-F^{\rm ref}_{i,\alpha}\Big| +
 \mathcal{L}_{\rm nh}
\label{eq:loss}
\end{equation} 
using AMSGrad\cite{reddi2018convergence}. Here, $E_{\rm ref}$ and
$F^{\rm ref}_{i,\alpha}$ are the reference energy and the force
components for atom $i$ with $(\alpha = x, y, z)$.  Parameters $w_{\rm
  E}$ and $w_{\rm F}$ are the weighting parameters (hyperparameters)
determining the relative contribution of individual error terms to the
loss function and $\mathcal{L}_{\rm nh}$ denotes a
``nonhierarchicality penalty'' which adds a penalty if the predictions
of the individual modules do not decay with increasing depth in the
PhysNet architecture.\\
  
\noindent
Six different dataset sizes were considered for training on clean and
perturbed energies together with clean and perturbed forces for
H$_2$CO. For training on energies-only the training dataset sizes were
$N_{\rm train} = 500$, to 3000 in steps of 500. For each case the
validation dataset size is $\sim 1/8$ of the training dataset size
(ranging from $N_{\rm valid} = 63$ to 375), respectively. The
remaining structures from a total dataset size of 3601 constituted the
test datasets. Repeat trainings were carried out by maintaining the
hyperparameters, training and validation dataset sizes, but varying
the initialization.\\

\noindent
Learning curves\cite{cortes:1993,muller:1996} were generated for all
clean and perturbed datasets by computing the dataset size dependent
root-mean squared errors (RMSEs) for the predicted energy (RMSE$(E)$)
or force (RMSE$(F)$). In such curves, the lowest and/or the average
RMSE of the two training runs for each dataset size are
reported. Training of a particular model continued until the
performance on the validation set did not improve anymore as judged by
the loss, see Eq. \ref{eq:loss}, or loss on the validation set started
to increase again due to overfitting. Typically, the model with the
lower RMSE$(E)$ was further analyzed.\\

\subsection{Electronic Structure Calculations}
All electronic structure calculations in the present work were carried
out using MOLPRO\cite{molpro} or Gaussian09.\cite{g09} The energies
and forces for H$_2$CO at the 3601 geometries obtained from normal
mode sampling\cite{smith2017ani} have been determined
previously.\cite{MM.tl:2021} To quantify the uncertainties incurred by
converging the Hartree-Fock wavefunction to different thresholds,
calculations at the MP2/aug-cc-pVTZ level of theory using the
Gaussian09 software package were carried out. The density based
convergence limits considered in the SCF were $10^{-8}$, $10^{-6}$ and
$10^{-4}$. In Gaussian09 the default convergence criterion on the RMS
density matrix is $10^{-8}$.\\

\noindent
For HONO, energies and forces were calculated at the MP2/aug-cc-pVTZ
level of theory for $6406$ geometries using MOLPRO\cite{molpro} and
were previously
published.\cite{MM.tl:2021,silvan_kaser_2021_4585449}. The dataset
contains geometries from normal mode sampling at temperatures ranging
from $T = 10$ to $2000~K$, from $NVT$ simulations run at $1000~K$
using the semiempirical GFN2-xTB method\cite{bannwarth2019gfn2} and
geometries along particular normal modes. Electronic structure
calculations for HONO at the coupled-cluster level of theory reveal
that the molecule possesses multi-reference character for various
structures as indicated by analyzing their $T_1$
characteristic\cite{lee1989diagnostic} determined from CCSD(T)
calculations. Additionally, energies for HONO were determined at the
MRCI/aug-cc-pVTZ level following initial complete active space (CAS)
calculations with CASSCF(16,12) for a total of 5987 structures. In
this case no analytical forces are available (see below).\\

\section{Results}
In the following, results for training with clean and perturbed
datasets of different sizes are reported. First, the modelled noise is
put into perspective by comparing with \textit{ab initio} calculations
employing different SCF convergence thresholds. Next, the effect of
perturbed energies on the learning is assessed for H$_2$CO, followed
by examining the effect of adding perturbations to the forces. Then,
the role of the hyperparameters is considered. Finally, learning in
the context of electronic structure calculations is probed by
generating datasets from energies based on reference calculations
converged to different thresholds in the SCF procedure and by
considering multi-reference effects for HONO.\\

\subsection{The Modelled Noise}
Quantum chemical calculations at all levels of theory require
convergence of certain quantities to within given thresholds. This
leads to aleatoric errors that introduce uncertainties
in the computed quantities such as energies or forces. 
For a set of nuclear geometries on which
the PES is trained, all reference calculations are ideally guaranteed
to have converged to at least within the required convergence
threshold. However, for individual geometries there are still
differences in the actual value to which convergence has been
reached. For example, for H$_2$CO energies at the MP2/aug-cc-pVTZ
level of theory calculated with Gaussian09 the achieved level of
convergence ranges from $0.37\times10^{-6}$ to $0.94\times10^{-6}$ for
a threshold of $10^{-6}$ applied to the RMS density matrix.\\

\begin{figure}[h]
\begin{center}
\includegraphics*[scale=0.65]{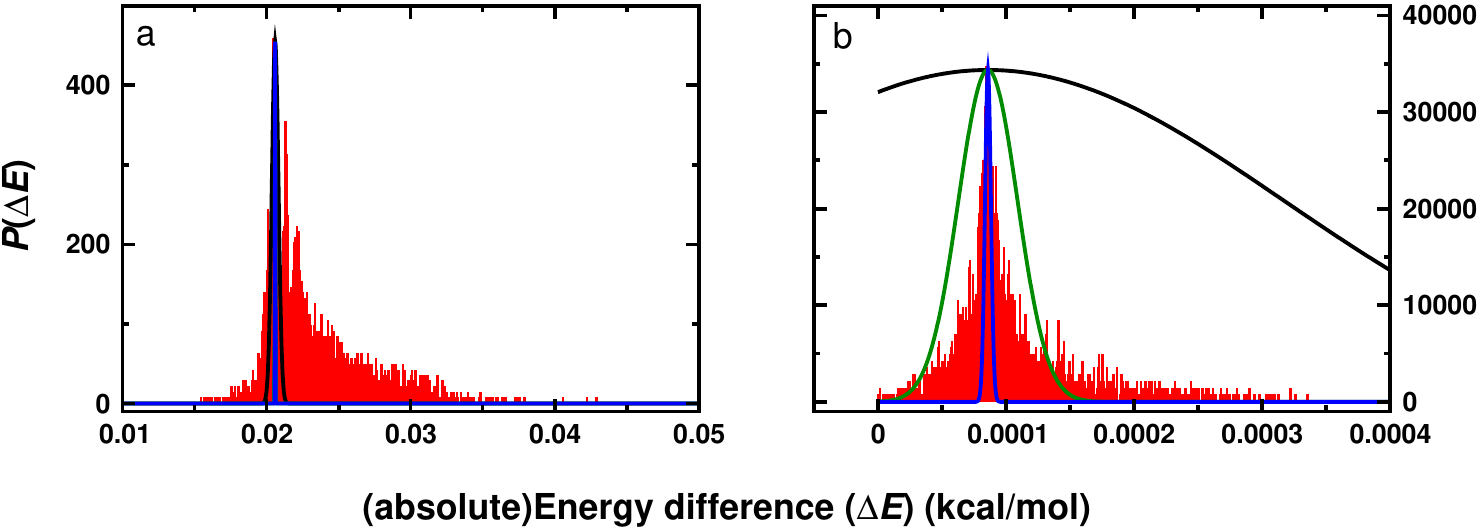}
\caption{Normalized probability distribution (red histograms) of the
  absolute energy differences $P(\Delta E)$ between the MP2 energies
  for H$_{2}$CO for $3601$ geometries with three ($10^{-4}$, $10^{-6}$
  and $10^{-8}$) different convergence limits in the SCF. The $\Delta
  E$ between energies calculated with SCF convergence limits of
  $10^{-4}$ and $10^{-6}$ is shown in panel a. Panel b shows the
  result for $10^{-6}$ and $10^{-8}$ convergence limits. Both panels
  also report three Gaussian curves with peak positions fixed at the
  peak of $P(\Delta E)$ and standard deviations (SDs) of $2.3\times
  10^{-4}$ (black line), $2.3\times 10^{-5}$ (green line) and
  $2.3\times 10^{-6}$(kcal/mol) (blue line). These SDs correspond to
  those used to generate random Gaussian noise that was added to the
  clean H$_{2}$CO/RKHS energies and forces.}
 \label{fig:fig1}
\end{center}  
\end{figure}

\noindent
To put the Gaussian-modelled noise introduced in the methods section
and uncertainties in energies arising in quantum chemical calculations
into perspective additional \textit{ab initio} calculations for the
3601 geometries of H$_2$CO used in previous work were carried
out.\cite{MM.tl:2021} All energies were recomputed at the
MP2/aug-cc-pVTZ level of theory after converging the SCF wavefunction
to within different prescribed convergence criteria. The distribution
of energy differences for convergence $10^{-4}$ and $10^{-6}$ is
reported in Figure \ref{fig:fig1}a and Figure \ref{fig:fig1}b shows
the distribution of energy differences with convergence $10^{-6}$ and
$10^{-8}$. The distributions are asymmetric with maxima at $\sim 0.02$
kcal/mol and $\sim 0.0001$ kcal/mol, respectively.\\

\noindent
Superimposed on the distributions in Figures \ref{fig:fig1}a and b are
Gaussian distributions centered around the mean with widths $10^{-7}$
(blue), $10^{-6}$ (green) and $10^{-5}$ (black) eV ($\sim 2.3 \times
10^{-6}$ to $2.3 \times 10^{-4}$ kcal/mol). These correspond to the
standard deviations of the generated Gaussian noise. Default
convergence criteria used in Gaussian09 are expected to be applicable
to typical organic molecules, whereas more lenient convergence
criteria of $10^{-4}$ to $10^{-5}$ on the density need typically be
used for molecules with more challenging electronic structure, such as
metal-containing systems.\cite{MM.fes:2004} For H$_2$CO using such
convergence thresholds lead to a rather wide and asymmetric
distribution $P(\Delta E)$ with a long tail towards larger $\Delta
E$. It is conceivable that for larger systems, such as [3Fe4S]
clusters,\cite{MM.fes:2004} the distributions $P(\Delta E)$ are at
least as broad as the one reported in Figure \ref{fig:fig1}a. For more
stringent convergence thresholds ($10^{-6}$ and $10^{-8}$) the
distribution $P(\Delta E)$ is reported in Figure \ref{fig:fig1}b. This
shows that for H$_2$CO drawing from a Gaussian with ${\rm SD} =
10^{-6}$ eV is representative, see red histogram and green Gaussian
distribution. The more pronounced tail in Figure \ref{fig:fig1}a may
imply that using less stringent convergence criteria for the
HF-wavefunctions $P(\Delta E)$ is less stochastic whereas for more
tightly converged calculations $P(\Delta E)$ is more similar to a
Gaussian distribution and the perturbations are more stochastic. It is
also noted that with tighter convergence criteria on the density the
maximum error shifts to smaller energy differences which is
accompanied by narrowing the distribution $P(\Delta E)$. The actual
magnitude of the aleatoric uncertainty and the shapes of $P(\Delta E)$
depend on the level of theory used.\\

\begin{figure}[h]
\begin{center}
\includegraphics*[scale=0.6]{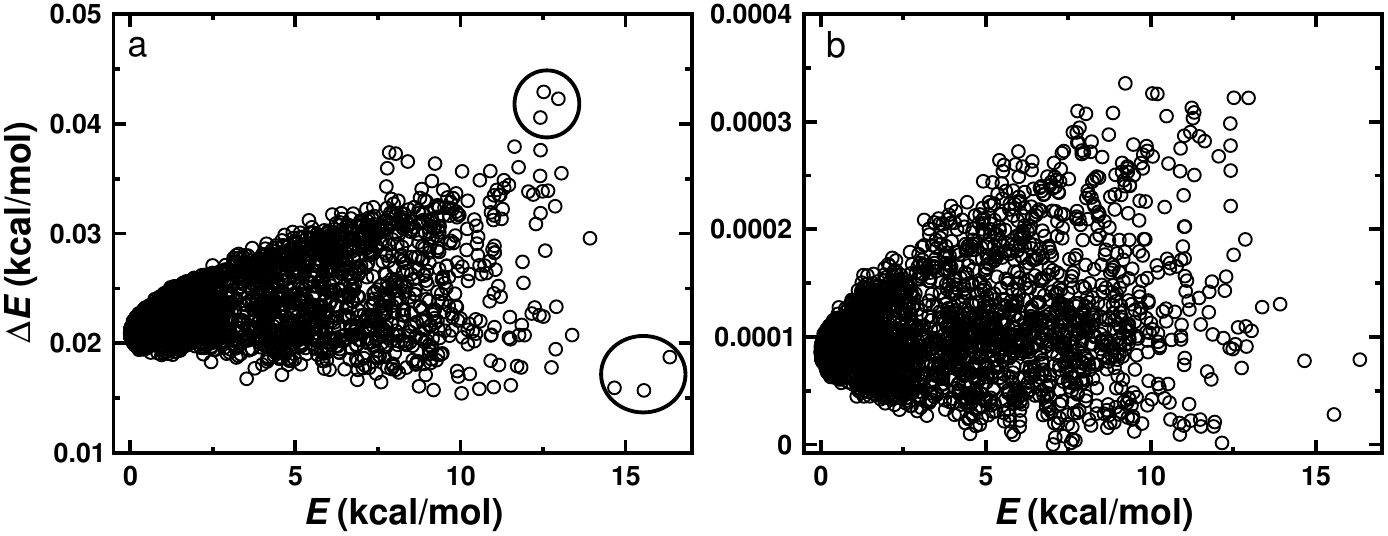}
\caption{MP2 energy difference $\Delta E$ between single point
  calculations for H$_2$CO with SCF-convergence $10^{-4}$ and
  $10^{-6}$ (panel a) and SCF-convergence $10^{-6}$ and $10^{-8}$
  (panel b). The $x-$axis in both panels are energies from
  calculations with SCF-convergence $10^{-6}$. The black circles
  mark structures with the highest $\Delta E$ and the highest energy
  and are illustrated in Figure~\ref{sifig:fig1}.}
\label{fig:fig2}
\end{center}
\end{figure}

\noindent
It is also found that for higher-energy structures the energy
differences from single-point calculations using the two different
convergence thresholds increases, see Figures~\ref{fig:fig2}a and
b. While a general trend suggests that high energy structures tend to
exhibit higher uncertainty ($\Delta E$), it is noteworthy that the
three structures possessing the highest energies have low $\Delta E$:
the three structures with the highest $E$ and $\Delta E$ (encircled in
Figure~\ref{fig:fig2}a) are shown in Figure~\ref{sifig:fig1}.\\

\begin{figure}[h]
\begin{center}
\includegraphics*[scale=0.7]{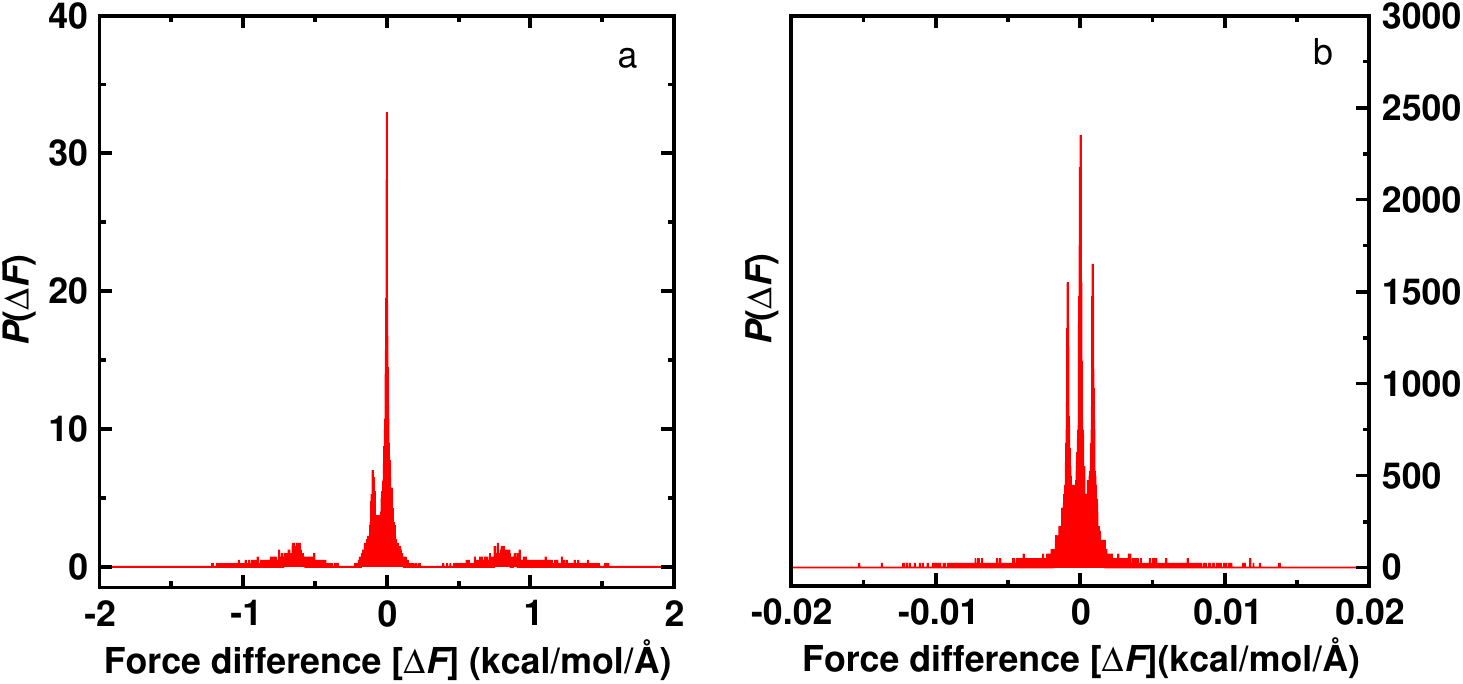}
\caption{The probability distributions of the differences in force
  between the MP2 forces, acting on all four atoms, of the H$_{2}$CO
  molecule calculated at 3601 geometries with three ($10^{-4}$,
  $10^{-6}$ and $10^{-8}$ ) different convergence limits in the
  SCF. The total force differences from calculations with convergence
  limits 10$^{-4}$ and $10^{-8}$ are shown in panel a, and panel b
  reports the results for convergence limits $10^{-6}$ and
  $10^{-8}$. The probability distributions of (absolute) force
  differences per atom type (C, O, H) are shown in Figure
  \ref{sifig:fig7}.}
\label{fig:fig8}
\end{center}  
\end{figure}

\noindent
The accuracy of PhysNet is estimated to be on the order of $\sim
10^{-3}$ kcal/mol for the present system and data set size. This
compares with noise levels of several 0.01 (Figure~\ref{fig:fig2}a)
and 0.0001~kcal/mol (Figure~\ref{fig:fig2}b) from converging the SCF
to thresholds between $10^{-4}$ and $10^{-8}$. Hence, while energy
fluctuations of 0.01~kcal/mol are somewhat above the accuracy of
current ML approaches, achieving a model accuracy on the order of
$10^{-4}$~kcal/mol poses a significant challenge\cite{MM.h2co:2020}
for PESs trained from statistical models.\\

\noindent
Similar to the analysis above, threshold-dependent differences can
also be obtained for the force components acting on the atoms. The
probability distributions of the difference for the $x-$, $y-$, and
$z-$components of the forces acting on each of the four atoms for all
3601 geometries of H$_{2}$CO with three different convergence limits
($10^{-4}$, $10^{-6}$ and $10^{-8}$) are shown in Figures
\ref{fig:fig8}a and b. As for the energies, these forces were also
calculated at the MP2/aug-cc-pVTZ level of theory, using the Gaussian
software package.\cite{g09} Although the peak of the distribution is
centered around $\sim 0.0$, non-negligible differences are found to
either side. Therefore, depending on the convergence criteria used at
the SCF step also influences the forces. Furthermore, SD=$10^{-5}$
eV/\AA\/ ($2.3\times 10^{-4}$ kcal/mol/\AA\/) and SD=$10^{-6}$
eV/\AA\/ ($2.3 \times 10^{-5}$ kcal/mol/\AA\/) used to generate noise
on the clean forces fall within the distribution range shown in Figure
\ref{fig:fig8}b, but are significantly smaller.\\

\subsection{H$_2$CO: Learning with Perturbed Energies}
\begin{figure}[h]
    \begin{center}
    \includegraphics*[width=\textwidth]{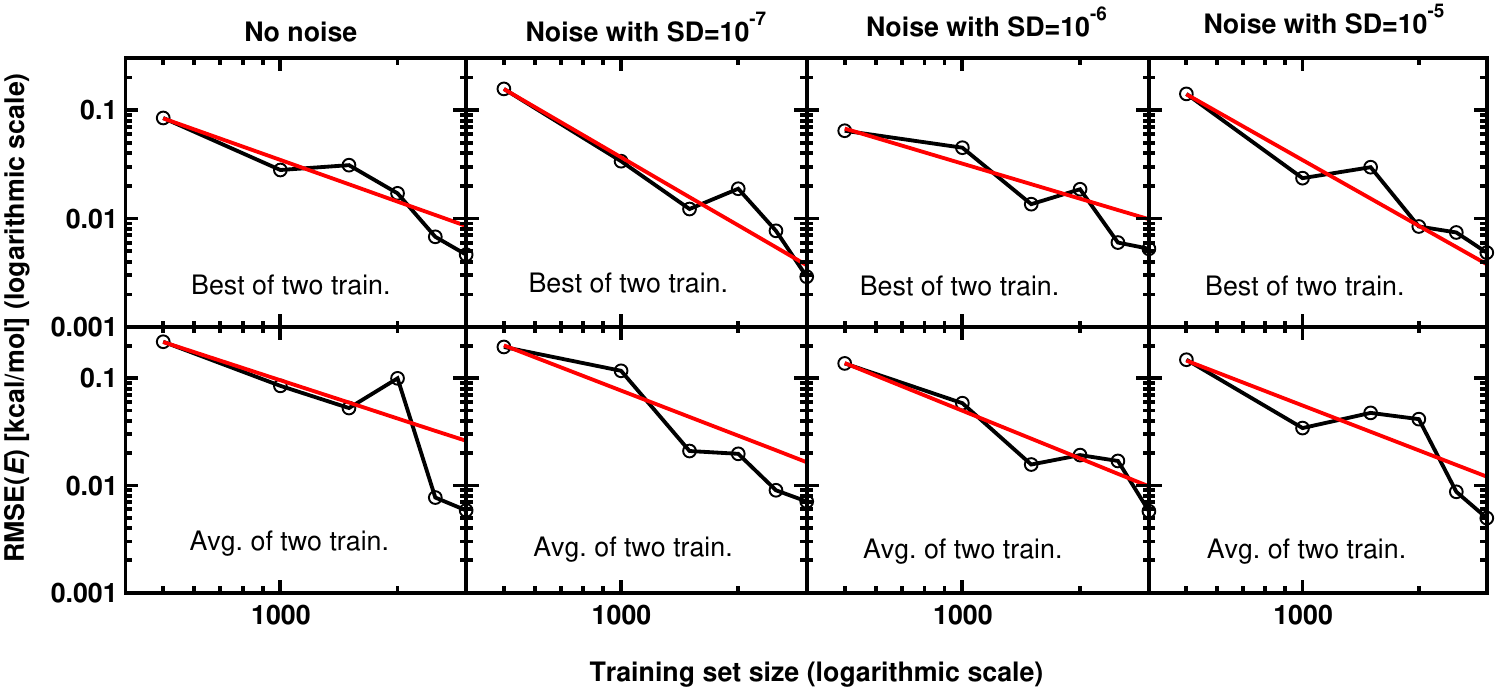}
    \caption{Energy learning curves (log-log plot) on the test data
      for the PhysNet based models for the H$_{2}$CO molecule. The
      reference energies are the clean RKHS energies and the perturbed
      RKHS energies with SD=10$^{-7}$ eV ($2.3\times 10^{-6}$
      kcal/mol), $10^{-6}$ eV ($2.3\times 10^{-5}$ kcal/mol) and
      $10^{-5}$ eV ($2.3\times 10^{-4}$ kcal/mol). In each case the
      models are trained on the same reference data but with dataset
      sizes of $N_{\rm train} = 500$, $1000$, $1500$, $2000$, $2500$
      and $3000$. Top row: performance of the model with the lower
      ${\rm RMSE}(E)$; bottom row: average of the two models.}
    \label{fig:fig3}
    \end{center}
\end{figure}

The learning curves for training PhysNet with ``energy-only'' for
H$_{2}$CO are shown in Figure~\ref{fig:fig3}. Considering training
using only energies (and no forces) is relevant for quantum chemical
methods that do not provide analytical gradients, such as multi
reference CI methods. Learning curves are reported for the clean
dataset and for perturbed energies with amplitudes of $10^{-5}$ eV
($2.3\times 10^{-4}$ kcal/mol), $10^{-6}$ eV ($2.3\times 10^{-5}$
kcal/mol), and $10^{-7}$ eV ($2.3\times 10^{-6}$ kcal/mol). Two
independent models were generated for each training dataset size and
results are reported for the model with the lower ${\rm RMSE}(E)$ (top
row) together with the average of the two (bottom row). Figure
\ref{fig:fig3} shows that the learning curves are hardly affected by
perturbed energies.\\

\noindent
Training based on ``energy-only'' data without and
with perturbation leads to improved models with increasing number of
training samples and the rate of learning (slope of red lines) is
unaffected, in particular for the averaged results. For the clean and
perturbed data sets and the different training set sizes $N_{\rm
  train}=500$ and $N_{\rm train} = 3000$ the [best, worst, average]
${\rm RMSE}(E)$ in kcal/mol range from [0.085; 0.354; 0.219] to
[0.005; 0.007; 0.006] (clean), [0.157; 0.234; 0.195] to [0.003; 0.011;
  0.007] (${\rm SD} = 10^{-7}$ eV), [0.065; 0.210; 0.138] to [0.005;
  0.006; 0.006] (${\rm SD} = 10^{-6}$ eV) and [0.141; 0.157; 0.149] to
[0.004; 0.005; 0.005] (${\rm SD} = 10^{-5}$ eV). Hence, for small
$N_{\rm train}$ addition of noise improves the average RMSE$(E)$
performance whereas for the largest training data set the performance
is virtually indistinguishable. One possible reason is that for the
present datasets the training hits the accuracy threshold of the
PhysNet architecture/representation which is estimated at $\sim
10^{-3}$~kcal/mol for the present dataset sizes and systems.
Therefore, the trained models are insensitive to noise with an
amplitude of $< 10^{-4}$~kcal/mol. Note that, on average (i.e. bottom
row of Figure~\ref{fig:fig3}), the model with the largest noise
performs better than the model without noise. This is likely to be
caused by the stochastic nature of NN based approaches and using an
ensemble with $N>2$ or, e.g., using a kernel based method will further
clarify the situation.\\

\begin{figure}[h]
    \begin{center} 
    \includegraphics*[scale=0.6]{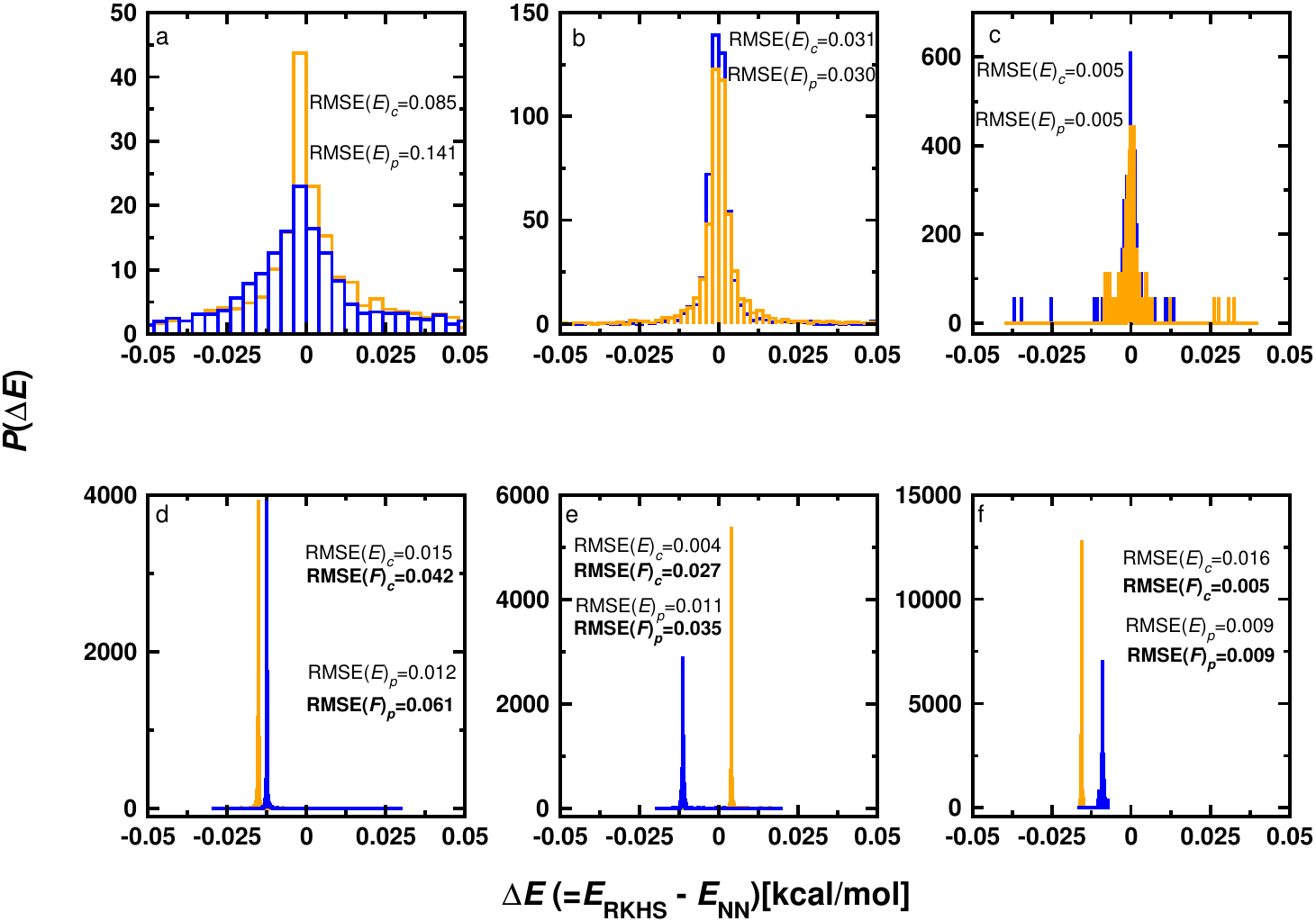}  
    \caption{The probability distribution of
      the energy prediction, $\Delta E = E_{\rm RKHS} - E_{\rm NN}$,
      for $N_{\rm train} = 500$ (a, d), 1500 (b, e), and 3000 (c,
      f). Panels a to c: ``energy-only'' training; perturbed energies
      contain random Gaussian noise with SD $=10^{-5}$ eV ($2.3\times
      10^{-4}$ kcal/mol). Panels d to f: clean energy + force
      training; perturbed forces contain random Gaussian noise with SD
      $=10^{-5}$ eV/\AA\/ ($2.3\times 10^{-4}$ kcal/mol/\AA\/) and
      the models are trained with force weighting hyperparameter
      $w_{\rm F} \sim 53$ (for trainings with $w_{\rm F} = 1$ see
      Figures \ref{sifig:fig6}a to c). For each dataset size, the best
      of the two trainings is shown. RMSE($E$)$_{c}$ and
      RMSE($E$)$_{p}$ represent the root-mean square error in energy
      for the clean (orange) and noisy (blue) datasets,
      respectively. RMSE($F$)$_{c}$ and RMSE($F$)$_{p}$ are for clean
      (orange) and noisy (blue) forces. See Figure \ref{sifig:fig2}
      for SD $=10^{-6}$ eV and SD $=10^{-6}$ eV/\AA\/ on energies and
      forces, respectively.}
    \label{fig:fig4}
    \end{center}
\end{figure}

\noindent
It is also of interest to consider energy differences between the
predicted $E_{\rm NN}$ and the reference energies $E_{\rm RKHS}$ for
the test structures. For learning on ``energies-only'', the
probability distributions of the energy difference, $\Delta E = E_{\rm
  RKHS} - E_{\rm NN}$, at the test points are shown in Figures
\ref{fig:fig4}a to c for $N_{\rm train} = 500$, 1500 and 3000 for
clean (orange) and perturbed (${\rm SD} = 10^{-5}$ eV ($2.3\times
10^{-4}$ kcal/mol), blue) data. While this noise level corresponds to
an extreme case and is likely not realistic for H$_2$CO, it is
representative for systems with intricate electronic structure. The
same assessment was also carried out for a noise level representative
of H$_2$CO (SD $=10^{-6}$ eV(/\AA) in Figure~\ref{sifig:fig2}a to c.\\

\noindent
For 500 training samples (Figure \ref{fig:fig4}a), the RMSE between
the clean and the perturbed energies for the best models differs by a
factor of $\sim 2$, i.e. 0.08 and 0.14 kcal/mol. For the worse of the
two models the RMSE($E$) is 0.35 and 0.16 kcal/mol for the clean and
perturbed data and illustrates the stochastic nature of training NNs.
The differences $\Delta E$ extend out to 0.5 kcal/mol for training on
the perturbed dataset and, using the best models, the probability to
find accurate predictions ($\Delta E \sim 0$ kcal/mol) is twice as
high for the clean dataset.  For a larger training set ($1500$
samples) the RMSE between reference and PhysNet energies decreases to
$0.03$ kcal/mol and is the same for clean and perturbed reference data
and for the largest training set size considered, the RMSE($E$) is
further reduced to 0.005~kcal/mol for training on both clean and noisy
data, see Figures \ref{fig:fig4}b and c. The maximum absolute $\Delta
E$ on the test set encountered for all training datasets using the
clean and perturbed data is comparable: 1.6 kcal/mol and 1.4 kcal/mol,
respectively.\\

\subsection{H$_2$CO: Learning with Perturbed Forces}
\begin{figure}[h]
\begin{center}
\includegraphics*[scale=0.5] {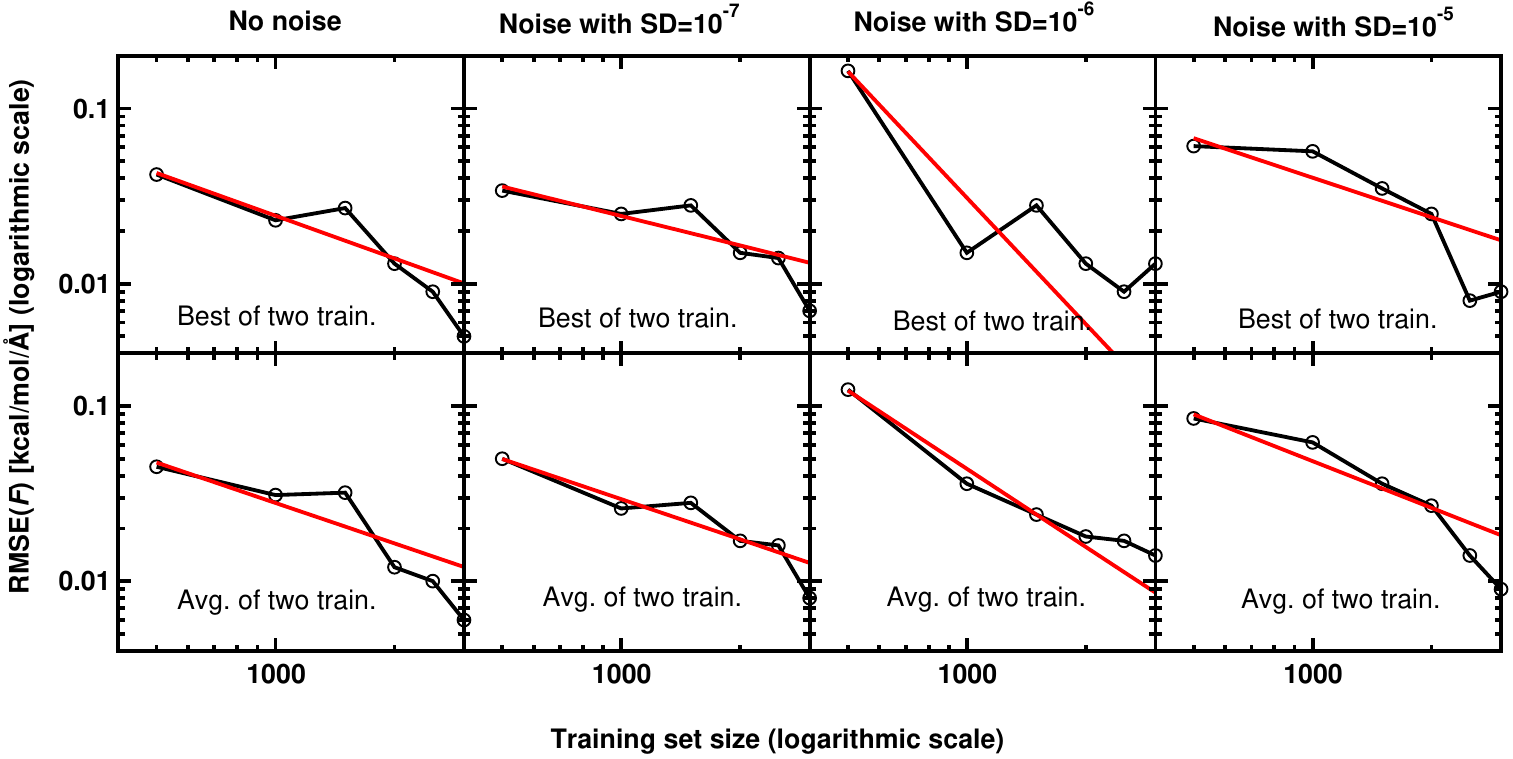}
\caption{Force learning curves (log-log plot) for PhysNet-based models
  for H$_{2}$CO. The reference forces are the clean RKHS forces (the
  clean dataset) and the RKHS forces with Gaussian noise of
  SD=$10^{-7}$ eV/\AA{} ($2.3\times 10^{-6}$ kcal/mol/\AA{}),
  $10^{-6}$ eV/\AA{} ($2.3\times 10^{-5}$ kcal/mol/\AA) and $10^{-5}$
  eV/\AA{} ($2.3\times 10^{-4}$ kcal/mol/\AA{}). In each case the
  models are trained on same reference data but with $500$, $1000$,
  $1500$, $2000$, $2500$ and $3000$ dataset sizes. Top row: model with
  lower RMSE$(E)$; bottom row: average over the two trainings. The
  energy is the clean RKHS energy and a $w_{\rm
    F} \sim 53$ is used (cf., Eq. (\ref{eq:loss})).}
\label{fig:fig5}
\end{center}
\end{figure}

For clean and perturbed forces, PhysNet was trained together with
clean energies. Noise was drawn from Gaussian distributions with zero
mean and with SD=$10^{-5}$ eV/\AA\/ ($2.3\times 10^{-4}$
kcal/mol/\AA\/) to $10^{-7}$ eV/\AA\/ ($2.3\times 10^{-6}$
kcal/mol/\AA\/) and the learning curves as log-log plots are shown in
Figure \ref{fig:fig5}. As for the energies, the training set size
varied between $500$ and $3000$ in steps of $500$ and two independent
PhysNet models were trained each. The top panels of Figure
\ref{fig:fig5} report results for the best of the two trained models
whereas the bottom row shows the average of the two independent
trainings. The noise levels employed here ($2.3 \times 10^{-4}$ to $2.3
\times 10^{-6}$ kcal/mol/\AA\/) are at the lower end of what was
observed (1 to 0.01 kcal/mol/\AA\/) from quantum chemical calculations
using the different convergence criteria, see Figure
\ref{fig:fig8}.\\

\noindent
Figure \ref{fig:fig5} indicates that training models with clean
energies and noisy forces can affect the learning curves for the
larger noise magnitude (SD = $10^{-6}$ eV/\AA\/ and $10^{-5}$
eV/\AA\/). This is unlike for the learning curves using energy-only
shown in Figure \ref{fig:fig3}. Therefore, perturbations on the forces
may affect the learning of PhysNet more significantly than introducing
perturbations on energies.\\

\noindent
Training on perturbed forces (with clean energies) yields the
probability distributions of the energy difference, $\Delta E = E_{\rm
  RKHS} - E_{\rm NN}$, reported in Figure \ref{fig:fig4}d to f. Here,
the spread of $P(\Delta E)$ is invariably smaller than for training on
``energy-only'' (compare with Figure \ref{fig:fig4}a to c) but the
centroid is shifted slightly away from zero by variable amounts
([$-0.015$ and $-0.012$], [$0.004$ and $-0.010$], [$-0.016$ and
  $\-0.008$] kcal/mol). Notably, the energy shift is larger for
training on clean data (orange) in two out of three cases, see
Figure~\ref{fig:fig4}d and f. In this context it is interesting to
note that for classification tasks it was found that including
measured amounts of noise can improve the generalization capabilities
of NN-based models.\cite{gardner:1989,benedetti:2023} Possible reasons
for the shifts away from $\Delta E = 0$ include i) the energy
distributions for training and test set do not center around the same
average or ii) training on energies and forces repositions the
centroid depending on the weighting $w_{\rm F}$ in Eq. \ref{eq:loss}
of the forces, compare Figures \ref{fig:fig4} and \ref{sifig:fig6}.\\

\noindent
Concerning point i) it is noted that the average reference energies on
the training and test data differ by 0.02, 0.07 and 0.20 kcal/mol for
$N_{\rm train}=500$, 1500 and 3000. These differences are exactly
matched by the trained NNs and compare with differences in the
centroids of up to 0.025 kcal/mol in Figure \ref{fig:fig4}d to
f. Consequently, differences between average training and test
energies seem an unlikely cause. As to point ii) it is found that the
centroids for $\Delta E$ are at zero for force hyperparameter $w_F =
1$, as will be discussed in more detail below. Hence, the offsets in
the energy prediction observed in Figure \ref{fig:fig4}d to f are most
likely due to the heavy weight on forces ($w_F \sim 53$) in the
training and the fact that reference energies and forces are not fully
consistent with one another. Results for training on perturbed data
with smaller noise level ${\rm SD} = 10^{-6}$ eV/\AA\/ ($2.3\times
10^{-5}$ kcal/mol/\AA\/) are reported in Figure~\ref{sifig:fig2}.\\

\noindent
The RMSE($F$) for training with perturbed forces (SD $=10^{-5}$
eV/\AA\/) and clean energies increases with decreasing size of the
training data set: {${\rm RMSE}(F)_{\rm c} = 0.005$ and ${\rm
    RMSE}(F)_{\rm p} = 0.009$ kcal/mol/\AA\/ for $N_{\rm train} =
  3000$; $0.027$ and $0.035$ kcal/mol/\AA\/ for $N_{\rm train} =
  1500$; $0.042$ and $0.061$ kcal/mol/\AA\/ for $N_{\rm train} = 500$,
  see Figures \ref{fig:fig4}d-f. These compare to ${\rm RMSE}(F)_{\rm
    c} = 0.007$ and ${\rm RMSE}(F)_{\rm p} = 0.009$ kcal/mol/\AA\/ for
  $N_{\rm train} = 3000$; $0.037$ and $0.037$ kcal/mol/\AA\/ for
  $N_{\rm train} = 1500$; $0.048$ and $0.109$ kcal/mol/\AA\/ for
  $N_{\rm train} = 500$ for the inferior models.  Thus, training on
  clean data yields RMSE($F$) which are lower throughout.  On the
  other hand, the trend is less clear for the energies, for which,
  e.g., RMSE($E$)$_c = 0.016$ and RMSE($E$)$_p = 0.009$ for a dataset
  size of 3000. Thus, perturbing the forces of H$_2$CO with
  large-amplitude artificial Gaussian noise can impact the
  learnability and quality of NN-based PESs. This is a particular
  challenge when NNs are trained from methods without analytical
  derivatives, such as MRCI+Q.  It should also be noted that from a
  practical perspective clean energies with perturbed forces is a
  less-likely scenario as perturbed forces are typically expected to
  originate from perturbations on the energies. On the other hand,
  forces at the CCSD(T)-F12 level have been found to be less accurate
  than machine precision\cite{werner:2018} and lead to ``floors'' in
  the learning curve\cite{MM.h2co:2020}. To more broadly corroborate
  these findings a larger number of NNs needs to be trained. It will
  also be of interest to consider alternative methods,
  e.g. kernel-based methods, which are known to reach lower out of
  sample errors that PhysNet.\cite{MM.h2co:2020}\\

\subsection{Probing Hyperparameters}
The choice of hyperparameters is known to affect training progress and
model quality for machine learned models, in particular
NNs.\cite{liao2022empirical,yuan2021systematic} This influence is
analyzed in the following by considering learning curves (see Figures
\ref{fig:fig6}a to d) for the PhysNet-based models resulting from
energy+force training for H$_2$CO using two different values for
weighting the force contribution, $\omega_F$, in the loss function,
see Eq. \ref{eq:loss}. The energies for training were clean RKHS
energies, whereas for the forces clean and Gaussian-perturbed forces
(largest amplitude ${\rm SD} = 10^{-5}$ eV/\AA\/ = $2.3\times 10^{-4}$
kcal/mol/\AA\/) were used. The two force weighting hyperparameters
were $w_{\rm F} \sim 53$ and $w_{\rm F} = 1$. Figures \ref{fig:fig6}a
and b show that for a large value of the hyperparameter $w_{\rm F}
\sim 53$ the error in the energy prediction does not decrease linearly
with increasing dataset size for both, clean and perturbed
forces. This changes if the value of the hyperparameter is decreased
to $w_{\rm F} = 1$, see Figures \ref{fig:fig6}c and d: without noise
the learning curve drops monotonically for almost all training data
set sizes (panel c) whereas with noise present first a plateau is
found after which a largely monotonous decrease follows (panel d).\\

\begin{figure}[h]
\begin{center}
\includegraphics*[scale=0.5]{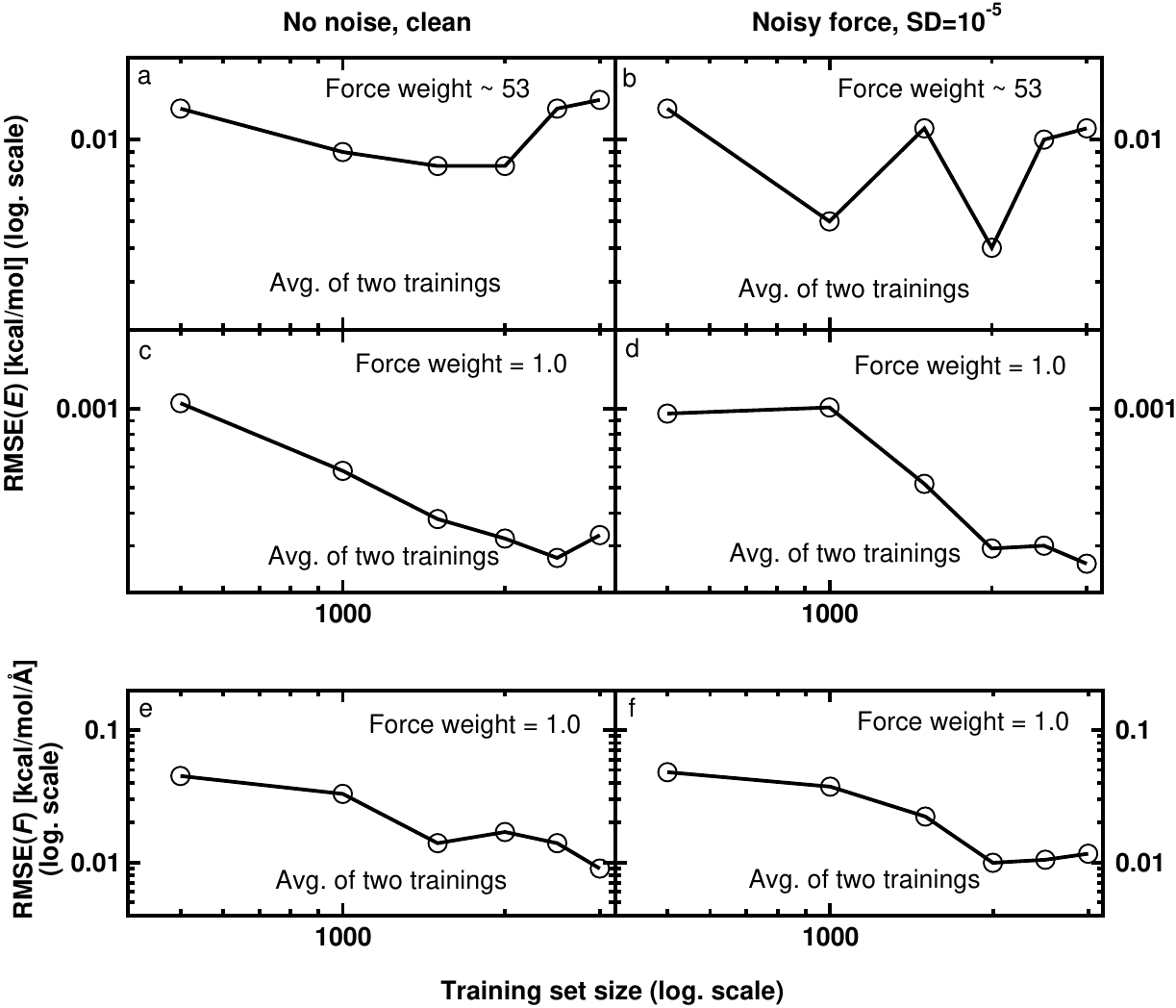}
\caption{Panels a-d: The energy learning curves for the PhysNet based
  models for the H$_{2}$CO molecule as log-log plot. The reference
  energies are the clean RKHS energies whereas the reference forces
  are the clean RKHS forces and the forces with Gaussian noise with
  SD=$10^{-5}$ eV/\AA{} ($2.3\times 10^{-4}$ kcal/mol/\AA{}). The
  used value of the weighting parameter of force in the loss function
  (cf., Eq. (17) of Ref. \citenum{MM.physnet:2019}) during training
  are $w_{\rm F} \sim 53$ (cf., panels a and b) and $w_{\rm F} = 1$
  (cf., panels c and d). Panels e and f: Force learning curves as
  shown in Figure \ref{fig:fig5} but only for the clean RKHS forces
  and forces with Gaussian noise with SD=$10^{-5}$ eV/\AA{}
  ($2.3\times 10^{-4}$ kcal/mol/\AA{}). The used value of the
  weighting parameter in the loss function during training is $w_{\rm
    F}=1$. The results presented in panels a-f are averaged over two
  independent trainings.}
\label{fig:fig6}
\end{center}
\end{figure}

\noindent
Corresponding force learning curves with a value of $w_{\rm F}=1$ for
the hyperparameter are shown in Figures \ref{fig:fig6}e and f. As
judged from the RMSE($F$) no effect from introducing noise on the
forces is visible and the performance of the two models is almost
identical. A comparison of these results with the curves in the bottom
row of Figure \ref{fig:fig5} reveals that the effect of the
hyperparameter value on learning the forces is not as severe as it is
for the energies. Furthermore, the probability distributions $P(\Delta
E)$ for the prediction error using $w_{\rm F} = 1$ do not exhibit
appreciable shifts away from $\Delta E = 0$, see Figure
\ref{sifig:fig6}, which differs from the findings for trainings with
$w_{\rm F} \sim 53$ as in Figure \ref{fig:fig4}. Hence, heavier
weighting of force components in the present case leads to small but
noticeable displacements of the average predicted energies. It is
noteworthy to mention that the H$_2$CO vibrational frequencies
calculated from the trained models with $w_{\rm F}=1$ and $w_{\rm F}
\sim 53$ only differ by a maximum of $0.2$ cm$^{-1}$, i.e. the quality
of the trained models only insignificantly depends on the choice of
hyperparameter.\\

\subsection{Perturbations from Multi-Reference Effects}  
Electronic structure calculation require certain input from the user
that influence the resulting uncertainties in the computed quantities,
such as energies or forces.  This includes, e.g., aleatoric
uncertainties introduced by convergence criteria on the SCF density
and epistemic errors due to smaller or larger basis sets used in the
calculations.  Furthermore, perturbations can also originate from
multi-reference effects. Recently, a detailed investigation of
isomerization and decomposition processes of HONO and HNO$_{2}$ has
been carried out.\cite{chen2019decomposition} State-of-the-art
electronic structure theory was used to compute the HNO$_{2}$ PES at
the CASPT2 and/or MRCI+Q levels for the dissociation pathways leading
to H+NO$_{2}$ and OH+NO. It was specifically noted that single
determinant methods are not reliable due to the presence of low-lying
excited states. Therefore, HONO was considered a suitable small
molecule to assess the influence of perturbations arising from
multi-reference effects in the input data for training PhysNet.\\

\begin{figure}[h]
 \begin{center}
  \includegraphics*[scale=0.5]{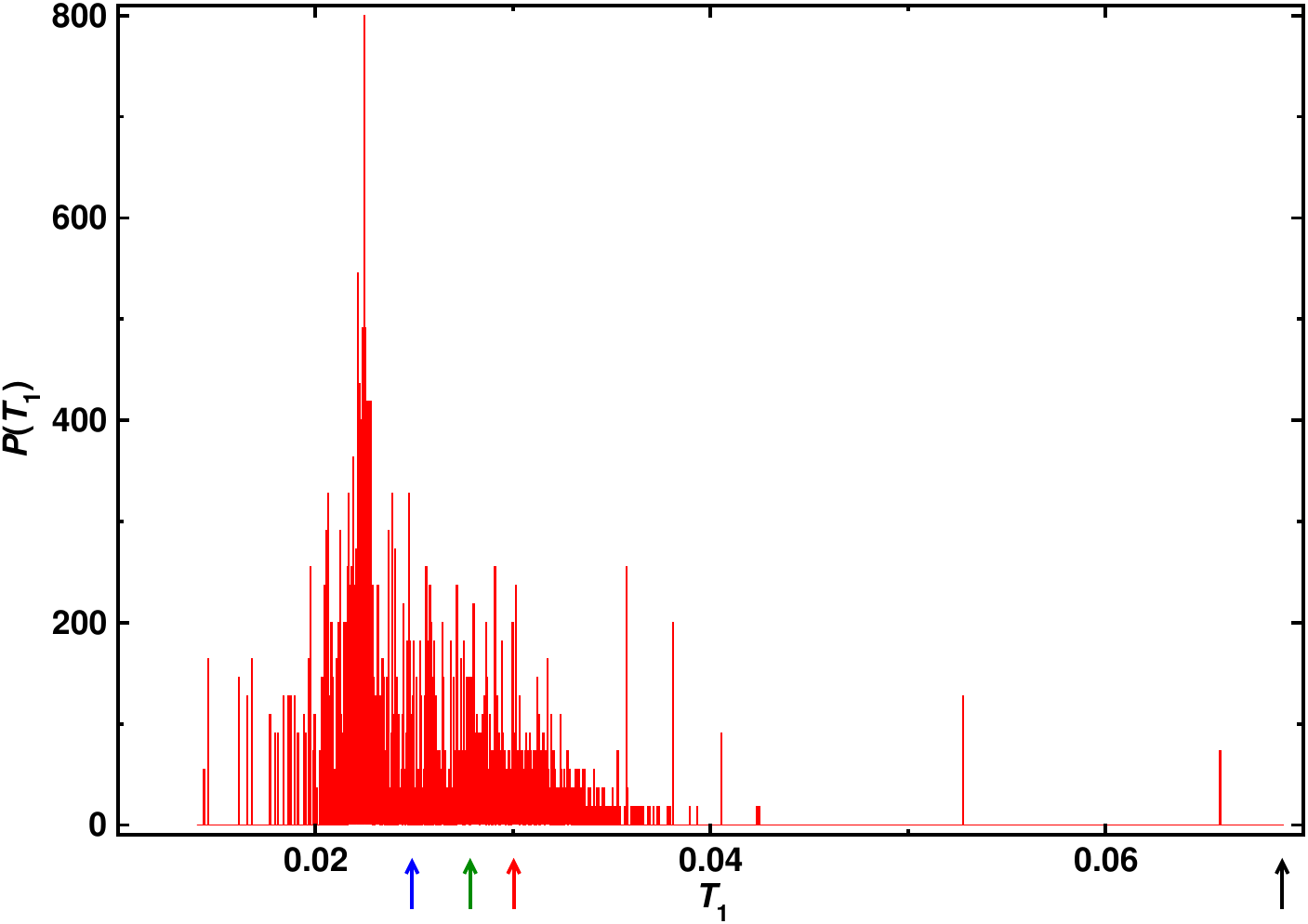}
\caption{The probability distribution of $T_{\rm 1}$ values of HONO
  from CCSD(T)/aug-cc-pVTZ calculations for $3375$ geometries. The
  maximum value of $T_{\rm 1}$ found in the considered $3375$, $2875$,
  $2375$ and $1875$ structures are $\sim 0.070$, $\sim 0.030$, $\sim
  0.028$, $\sim 0.025$, respectively. These are indicated here by
  black, red, green and blue arrows. The broad distribution is akin to
  an aleatoric uncertainty.}
\label{fig:fig9}
\end{center}
\end{figure}

\noindent
In the following, the $T_{\rm 1}$ diagnostic was used to quantify the
amount of nondynamic electron correlation and whether or not employing
a single-reference treatment is appropriate.\cite{lee1989diagnostic}
Typically, a value of $T_{\rm 1} > 0.02$ is taken as an indication
that a multi-reference approach should be employed for a particular
geometry.\cite{lee1989diagnostic} Alternatively, the $D_1$ diagnostic
could also be used which was, however, not done in the present
work.\cite{janssen:1998} In addition to the electronic structure
calculations of HONO at the MP2/aug-cc-pVTZ level (see Methods),
CCSD(T)/aug-cc-pVTZ calculations were carried out to obtain the
$T_{\rm 1}$ amplitudes. The distribution of $T_{\rm 1}$ values (for
$3375$ HONO structures) reported in Figure \ref{fig:fig9} demonstrates
that a multi-reference treatment is required for 3235 ($\sim 96$ \%)
of these geometries if the criterion $T_{\rm 1} > 0.02$ is used.\\

\begin{figure}[h]
\begin{center}
  \includegraphics*[scale=0.65]{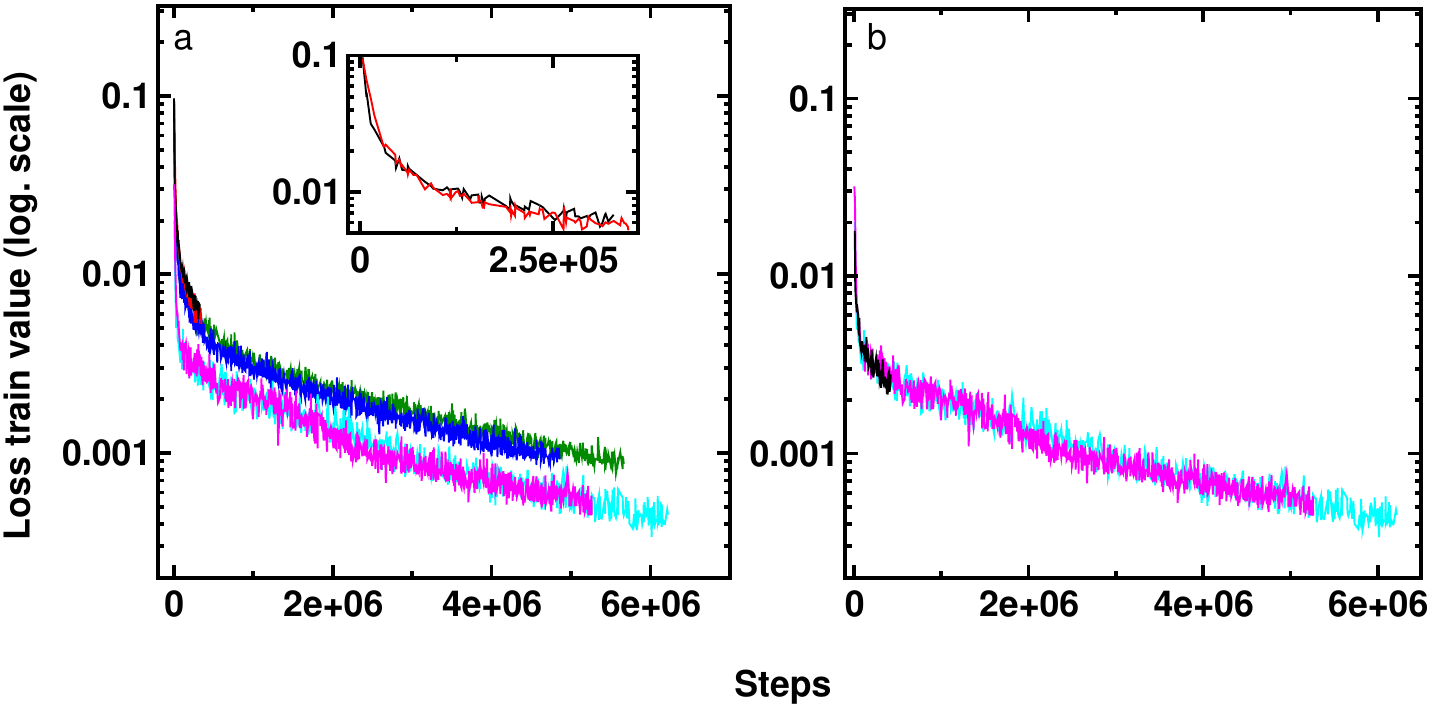}
\caption{The values of the loss function (training on energy-only) for
  HONO (panel a and inset) and H$_2$CO (panels a and b) using
  different MP2 training dataset sizes up to the point at which
  overfitting starts. Panel a: The loss function value for HONO from
  training datasets with 3000 (black, all), 2500 (red, $T_1 \leq
  0.030$), 2066 (green, $T_1 \leq 0.028$) and 1631 (blue, $T_1 \leq
  0.025$) structures. Panel b: The loss function value for H$_2$CO
  with 3000 (cyan), 1500 (magenta) and 500 (black) training dataset
  sizes. In panel a, results for training with 3000 (cyan) and 1500
  (magenta) reference structures for H$_2$CO are also shown for
  comparison. The inset in panel a is a close-up for HONO for 3000 and
  2500 training set sizes.}
\label{fig:fig10}
\end{center}
\end{figure}

\noindent
In the following it is of interest to determine whether training
PhysNet on subsets of the entire MP2 data set of energies a) displays
differences in the learning curves depending on how many energies with
multi-reference character (as per the $T_1$ amplitude) are retained
and b) how the model performance is influenced. To this end, the
training set for HONO was restricted to structures with progressively
lower $T_1$ amplitudes. This resulted in datasets with $2875$ ($T_1
\leq 0.03$), $2375$ ($T_1 \leq 0.028$), and $1875$ ($T_1 \leq 0.025$)
structures. The parts of the distribution $P(T_1)$ that were retained
for training are to the left of the colored arrows in Figure
\ref{fig:fig9}.\\

\noindent
Training for each of these datasets was carried out using 87 \%
training structures and 13 \% for validation, and the training loss is
shown up to the point for which convergence is reached, \textit{i.e.}
the validation loss does not decrease further or even increases again
due to overfitting. The value of the loss function for HONO
(calculated on the training data) is reported in Figure
\ref{fig:fig10}a. If structures with $T_1 \leq 0.03$ (red trace) are
retained, the loss is similar to the one for using the entire dataset
(black trace) without restriction on structures with particular $T_1$
amplitude, see also inset in Figure \ref{fig:fig10}a. The value of the
loss function reaches $\sim 0.006$ for these two training dataset
sizes after $\sim 3 \times 10^5$ steps. However, for progressively
tighter criteria on structures with lower $T_1$ amplitudes (green and
blue traces), training continues for at least one order of magnitude
more steps ($\sim 5 \times 10^6$) and reaches lower values for the
loss function by about a factor of 5.\\

\noindent
Similarly, training on MP2 energies for H$_2$CO using $500$, $1500$,
and $3000$ training structures were carried out and the value of the
loss function is reported in Figure \ref{fig:fig10}b (black, magenta,
cyan). Here it is found that with the smallest training set the loss
function improves for about one order of magnitude fewer steps
compared with the two larger training set sizes. This differs
for HONO in that the largest training set size was found
to stop learning first. For direct comparison
with HONO, the magenta and cyan traces (H$_2$CO for 1500 and 3000
training data from panel b) are also reported in Figure
\ref{fig:fig10}a. This demonstrates that by eliminating structures
with high $T_1$ from the HONO data set but retaining a sufficiently
large training data set size ($\geq 1500$ structures) the loss
function decays in a comparable fashion as for the H$_2$CO MP2
energies which do not contain potential noise from multi-reference
calculations. Hence, the presence of energies for structures with
multi-reference character, but treated with a single-reference method,
compromises the learning capabilities of a model such as
PhysNet. Overall, however, the rate at which the loss function decays
and the best achievable model are inferior to training on a dataset of
clean, single-reference energies such as for H$_2$CO. One possibility
is that - as is the case for HONO - the mixture of energies that are
or are not affected by multi-reference effects, as measured by the
$T_1$ amplitude, leads to difficulties in the learning, which is
considered next.\\

\begin{figure}
\begin{center}
  \includegraphics*[width=\textwidth]{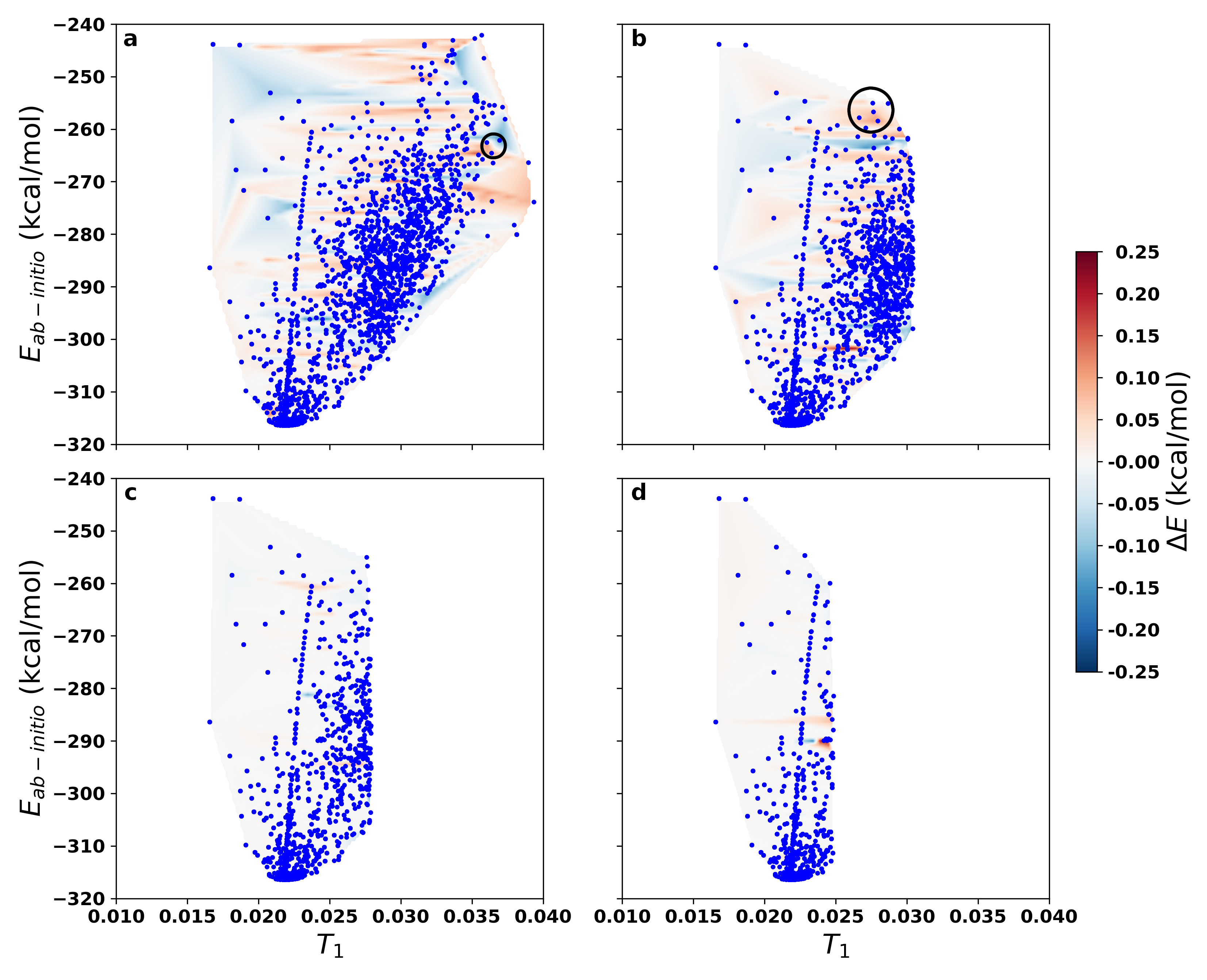}
\caption{Heatmap for the error distribution $\Delta E = E_{\rm ref} -
  E_{\rm NN}$ for HONO trained on MP2 energies depending on the $T_1$
  amplitude (from CCSD(T)/aug-cc-pVTZ calculations) and the total
  energy of a structure $E_{\rm ab-initio}$. Panel a: without
  restriction on the $T_1$ amplitude; panel b: $T_1 \leq 0.030$; panel
  c: $T_1 \leq 0.028$; panel d: $T_1 \leq 0.025$. Dark red and blue
  regions only occur if structures with $T_1 \geq 0.028$ are retained
  in the training. Structures for the encircled data points in panels
  a and b are reported in Figure \ref{sifig:fig3}. For the probability
  distributions $p(\Delta E)$, see Figure \ref{sifig:fig5}. Note that
  single outliers with larger $|\Delta_E|$ are excluded for clarity.}
\label{fig:fig11}
\end{center}
\end{figure}

\noindent
For this it was assessed whether the total energy of a particular
structure, the prediction error, and $T_1$ are correlated. Again, the
MP2 data sets for HONO with progressively lower $T_1$ amplitudes was
considered. Figure \ref{fig:fig11}a reports the results from training
PhysNet using all MP2 reference data, i.e. without restriction on
$T_1$ values. Qualitatively it is found that large values of $T_1$ are
associated with high energy structures. The converse is not
necessarily true: there are also high-energy structures that do
exhibit only moderately large $T_1$ amplitudes. Large positive (red
density) and negative (blue density) prediction errors arise and are
typically associated with high-energy structures and/or large values
of $T_1$. These prediction errors are skewed towards $\Delta E > 0$
(red), see also inset in Figure \ref{sifig:fig5}. Restricting the
training set to structures with $T_1 \leq 0.030$ leads to the
performance reported in Figure \ref{fig:fig11}b. Large positive (dark
red) and negative (dark blue) errors for the energy still occur. For
the training in Figure \ref{fig:fig11}c only structures with $T_1 \leq
0.028$ were retained which considerably reduces large positive
prediction errors and no large negative errors are found anymore. The
correlation between the magnitude of $T_1$ and the magnitude of the
prediction error is further corroborated by training on structures
with yet smaller $T_1 \leq 0.025$, see Figure
\ref{fig:fig11}d. Consequently, the widths for the prediction errors
decrease considerably, see Figure \ref{sifig:fig5}. In summary, this
analysis points to a direct relationship between the magnitude of
$T_1$ and the prediction errors from models trained on corresponding
subsets. \\

\noindent
Structures with particularly high $T_1$ and large prediction errors
are encircled in Figures \ref{fig:fig11}a and b and the corresponding
geometries are reported in Figures \ref{sifig:fig3} and
\ref{sifig:fig4}. It is found that in all cases the geometries are
perturbed away from the HONO equilibrium geometry in a pronounced
fashion. For example, the structure with $116.4^{\circ}$ dihedral
angle (see Figure \ref{sifig:fig3}a) has a noticeably longer $R_{\rm
  O_{B}H}$ bond length and widely open $\theta_{\rm O_{A}NO_{B}}$; the
structure with $-72.7^{\circ}$ dihedral angle (see Figure
\ref{sifig:fig3}b) has longer bond lengths for all three bonds,
whereas the structure with $58.2^{\circ}$ dihedral angle has a narrow
$\theta_{\rm O_{A}NO_{B}}$ angle.\\

\noindent
Finally, it is also of interest to compare the final model performance
on the training and test sets for reference MP2 and MRCI+Q
calculations, see Table \ref{tab:tab1}, which addresses epistemic
errors. Here, it was found that the RMSE$(E)$ on the training set
increased by about an order of magnitude in going from MP2 to MRCI+Q
reference data, largely irrespective of the size of the training set
$N_{\rm train}$. This does not translate directly to model performance
on the test data. In this case the trained models differ rather by a
factor of 2 to 3 for MP2 vs. MRCI+Q reference data. Nevertheless,
training on the MRCI+Q reference calculations yields models with
higher representation errors in comparison to those trained on MP2
data. Conversely, training a single-reference problem such as H$_2$CO
on MP2 reference data leads to models that are more accurate by an
order of magnitude compared with a multi-reference system (HONO) using
the same single-reference method (MP2). Besides effects from
single-/multi-reference H$_2$CO is a molecule with higher symmetry and
the energy range of the employed dataset is smaller by a factor of
$\sim 5$ (energy range of $\sim 15$~kcal/mol for H$_2$CO vs.
80~kcal/mol for HONO), which presumably make the learning process
easier for H$_2$CO.\\

\begin{table}
  \caption{Summary of the "energy-only" trainings for H$_{2}$CO (MP2)
    and HONO (*MP2, **MRCI+Q) along with the root-mean squared errors
    for the energy (rounded to six decimal place) on the training
    (RMSE($E$)$_{\rm train}$) and test (RMSE($E$)$_{\rm test}$) data
    given in kcal/mol. $^{***}$Average of five independent trainings
    with different seeds.}
\begin{tabular}{lllllll}
\hline \hline
Compound & $    $ $N_{\rm tot}$ & $    $ $N_{\rm train}$ & $    $ $N_{\rm valid}$ & $    $ $N_{\rm test}$ & ${\rm RMSE}(E)_{\rm train}$ & ${\rm RMSE}(E)_{\rm test}$\\
\hline
$    $ H$_{2}$CO & $  $ $3601$ & $  $ $3000$ & $  $ $375$ & $  $ $226$ & $        $ $0.002858$ & $      $ $0.005601$\\
$    $ H$_{2}$CO & $  $ $3601$ & $  $ $1500$ & $  $ $188$ & $  $ $1913$ & $       $ $0.003789$ & $      $ $0.019181$\\
\hline
$    $ HONO* (full) & $  $ $6406$ & $  $ $3000$ & $  $ $375$ & $  $ $3031$ & $       $ $0.051401$ & $       $ $0.164302$\\
$    $ HONO* ($T_1 \leq 0.030$) & $  $ $2875$ & $  $ $2300$ & $  $ $288$ & $  $ $287$ & $       $ $0.025900^{***}$ & $       $ $0.233161^{***}$\\
$    $ HONO* ($T_1 \leq 0.030$) & $  $ $2875$ & $  $ $2500$ & $  $ $375$ & $  $ $0$     & $       $ $0.057809$            & $       $ n.a. \\
$    $ HONO* ($T_1 \leq 0.028$) & $  $ $2375$ & $  $ $1900$ & $  $ $238$ & $  $ $237$ & $       $ $0.020399^{***}$ & $       $ $0.291417^{***}$\\ 
$    $ HONO* ($T_1 \leq 0.028$) & $  $ $2375$ & $  $ $2066$ & $  $ $309$ & $  $ $0$ & $       $ $0.006552$ & $       $ n.a. \\
$    $ HONO* ($T_1 \leq 0.025$) & $  $ $1875$ & $  $ $1500$ & $  $ $188$ & $  $ $187$ & $       $ $0.016898^{***}$ & $       $ $0.290245^{***}$\\
$    $ HONO* ($T_1 \leq 0.025$) & $  $ $1875$ & $  $ $1631$ & $  $ $244$ & $  $ $0$ & $       $ $0.005003$ & $       $ n.a. \\
\hline
$    $ HONO** & $  $ $3150$ & $  $ $2520$ & $  $ $315$ & $  $ $315$ & $       $ $0.155482^{***} $ & $       $ $0.434823^{***}$\\
$    $ HONO** & $  $ $3150$ & $  $ $2205$ & $  $ $276$ & $  $ $669$ & $       $ $0.131824 $ & $       $ $0.358221$\\
$    $ HONO** & $  $ $3150$ & $  $ $1890$ & $  $ $236$ & $  $ $1024$ & $      $ $0.132721 $ & $       $ $0.579272$\\
$    $ HONO** & $  $ $3150$ & $  $ $1575$ & $  $ $197$ & $  $ $1378$ & $      $ $0.165319 $ & $       $ $0.716033$\\
\hline \hline 
\label{tab:tab1}
\end{tabular}
\end{table}

\section{Conclusions}
In this contribution the effect of perturbations in the reference data
obtained from quantum chemistry calculations used for training
ML-based potential energy surfaces is quantified. The work explores
the effects of a) convergence criteria in quantum chemical
calculations and b) using single-reference methods for systems with
multi-reference character on the \textit{ab initio} data and
subsequent learning. H$_{2}$CO is a small, symmetric and closed-shell
molecule for which high-quality reference calculations for structures
around the minimum geometry are standard. Changing the convergence
limits of the SCF calculations to those typical of larger and
electronically more demanding systems, such as metal-complexes, may
affect the trainability of PhysNet if perturbations are present in the
forces. Nevertheless, for perturbed forces the effects on the trained
NNs found in the present work were still comparatively small although
the magnitude of the perturbations used on the clean data was rather
small compared with what is found from quantum chemical calculations
employing difference convergence thresholds in the SCF
calculations. It is possible that with alternative ML strategies, such
as the kernel-based Faber/Christensen/Huang/von Lilienfeld (FCHL)
approach\cite{faber2018alchemical,christensen2020fchl}, lower
representation errors can be reached and the noise floor is detected
earlier and even "energy-only" learning is sensitive to perturbations
in the data.\\

\noindent
For HONO, which requires a multi-reference treatment, a clear
correlation between the accuracy of the final model and the $T_1$
amplitude - which is a measure of nondynamic correlation - was found,
see Figure \ref{fig:fig11}. Furthermore, the quality of the models as
judged from the RMSE$(E)$ on the training and test data set, clearly
differs between H$_2$CO and HONO at the MP2 levels of theory. These
effects are even more pronounced when comparing model performance for
HONO on MP2 and MRCI+Q reference data, see Table \ref{tab:tab1}.\\

\noindent
Within a broader scope, multi-reference CI and second order
perturbation theory (CASPT2) calculations require the definition of an
”active space” of molecular orbitals that are treated within the
multi-reference calculation. Usually, the orbitals that constitute the
active space are not uniquely defined.  For example, which orbitals to
retain in the active space and whether or not to introduce new
orbitals may well change for an ensemble of molecular
structures. Hence, it is expected that working with one definition of
the CAS for reactions -- an ensemble of structures including reactant,
product, and transition state(s) -- a union of orbitals specific for
each structure should be employed for a balanced treatment. This
necessarily increases the CAS and renders routine calculations for
larger molecules impractical. Furthermore, the simulation of excited
states for one of these structures would require yet another set of
orbitals for the active space. This freedom in choosing suitable
orbitals for an ensemble of structures introduces epistemic
uncertainty into such calculations whereas the fact that for a given
set of orbitals the weight of each Slater-determinant changes leads to
aleatoric uncertainty.\\

\noindent
It is of interest to note that far away from the minimum energy
structure and for electronically more demanding systems, e.g. those
containing metal centers or halogens, multi-reference effects are
rather the norm than the exception.\cite{wang:2015} Therefore,
training machine learning models - in particular neural networks as
was done in the present work - requires circumspection in assessing
the role of noise, both aleatoric and epistemic, when assessing model
performance. Finally, it is important to mention that further work is
required for a broader and more comprehensive understanding and
characterization of what role noise/errors in the input data plays for
the performance of NN-based PESs.\\

\section*{Acknowledgment}
Support by the Swiss National Science Foundation through grants
200021{\_}117810, 200021{\_}215088, the NCCR MUST (to MM), and the
University of Basel is acknowledged. This work was also supported by
the United State Department of the Air Force, which is gratefully
acknowledged (to MM). This project has received funding from European
Union’s Horizon 2020 under MCSA Grant No 801459, FP-RESOMUS. Valuable
correspondence on complete active spaces with Prof. R. Lindh is
acknowledged.

\section*{Data Availability Statement}
The clean and perturbed RKHS energies and all reference data
employed in the present study are available at
\url{https://github.com/MMunibas/noise}.\\

\bibliography{refs.tidy}

\clearpage

\renewcommand{\thetable}{S\arabic{table}}
\renewcommand{\thefigure}{S\arabic{figure}}
\renewcommand{\thesection}{S\arabic{section}}
\renewcommand{\d}{\text{d}}
\setcounter{figure}{0}  
\setcounter{section}{0}  

\noindent
{\bf Supporting Information: On the Effect of Aleatoric and
  Epistemic Errors on the Learnability and Quality of NN-based
  Potential Energy Surfaces}\\

\begin{figure}
 \begin{center}
     \includegraphics*[scale=0.65]{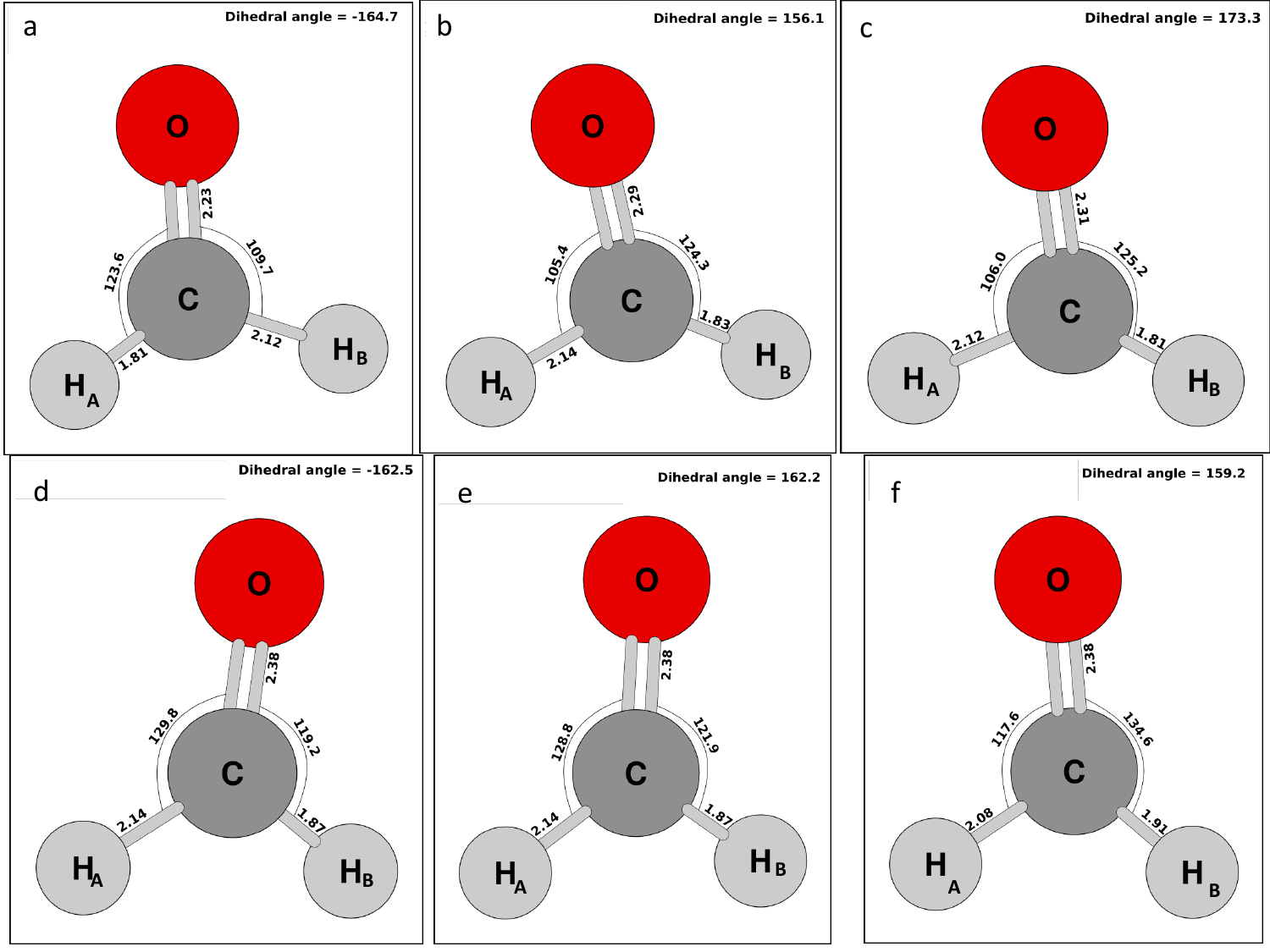}
     \caption{Structures for H$_2$CO corresponding to data points
       encirceled in Figure \ref{fig:fig2}a. Panels a to c: Structures
       for data points with $E > 14$ kcal/mol. Panels d to f:
       structures with $\Delta E > 0.04$ kcal/mol. The distances and
       angles are in a$_0$ and degree units, respectively. The
       equilibrium (MP2/aug-cc-pvtz) structure of H$_{2}$CO is planar
       (dihedral of $180^\circ$) with $R_{\rm CO}=2.29$ a$_0$, $R_{\rm
         CH}=2.08$ a$_0$, $\theta_{\rm OCH}=121.7^{\circ}$. Thus all
       these six structures are distorted from the equilibrium
       geometry. Apart from the dihedral angle, the distortion for the
       structures in panels a to c primarily occurs due to shortening
       of one of the $\rm {CH}$ bonds and narrowing of one of the
       $\theta_{\rm OCH}$ angles. On the other hand, distortion for
       the bottom row (d-f) structures primarily takes place because
       of shortening of one of the $\rm {CH}$ bonds and opening up of
       one of the $\theta_{\rm OCH}$ angles.}
     \label{sifig:fig1}
 \end{center}
\end{figure}

\begin{figure}
    \begin{center}
    \includegraphics*[scale=0.7]{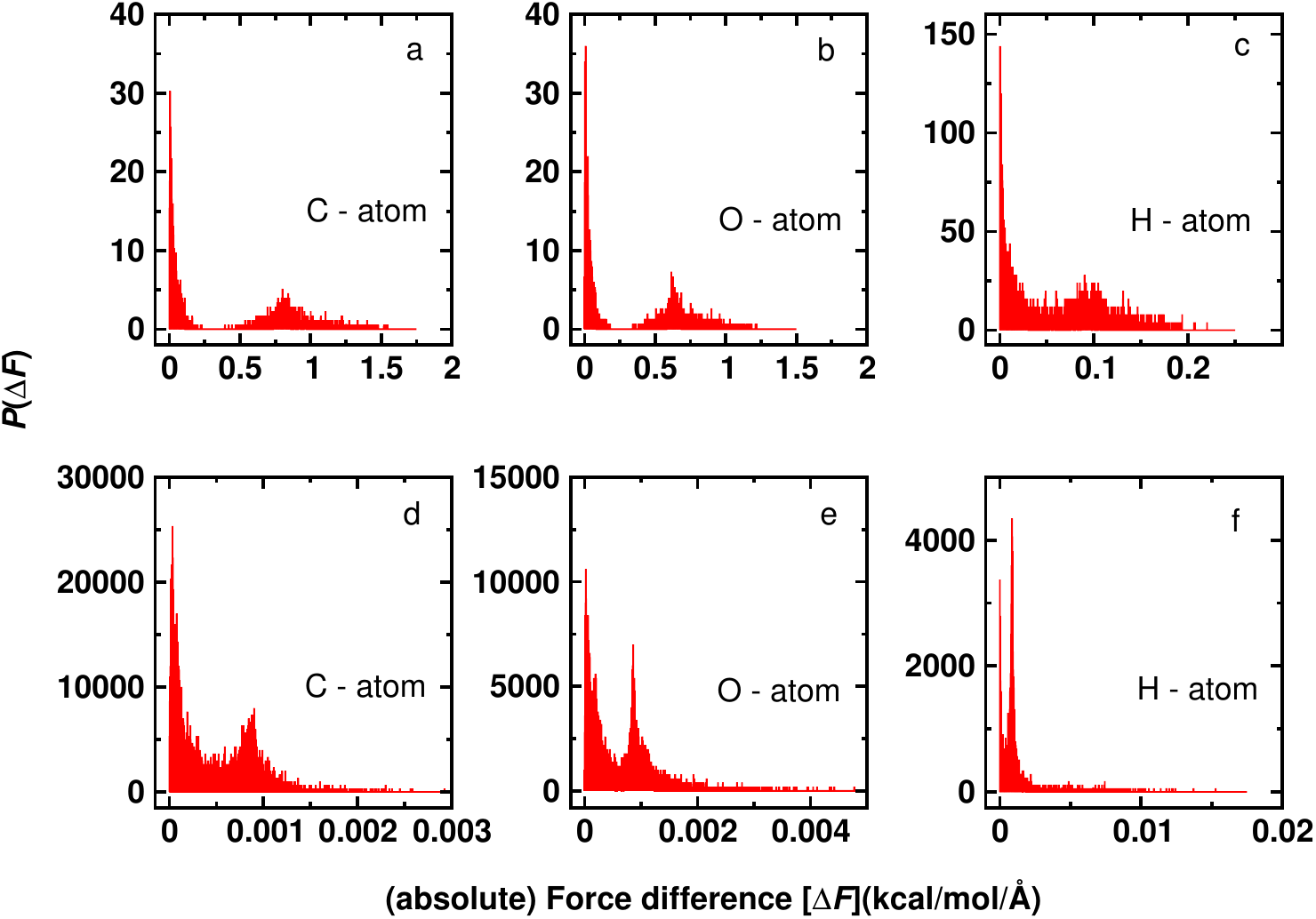}
    \caption{The probability distributions of (absolute) differences
      in force between the MP2 forces, acting on all the components of
      Carbon, Oxygen and one Hydrogen atom of the H$_{2}$CO molecule
      calculated at $3601$ geometries with three ($10^{-4}$, $10^{-6}$
      and $10^{-8}$) different convergence limits in the SCF. The
      results of $10^{-4}$ and $10^{-8}$ convergence limits are shown
      in panels a-c, whereas, panels d-f show the results for
      $10^{-6}$ and $10^{-8}$ convergence limits.}
    \label{sifig:fig7}
    \end{center}
\end{figure}

\begin{figure}
    \begin{center}
    \includegraphics*[scale=0.7]{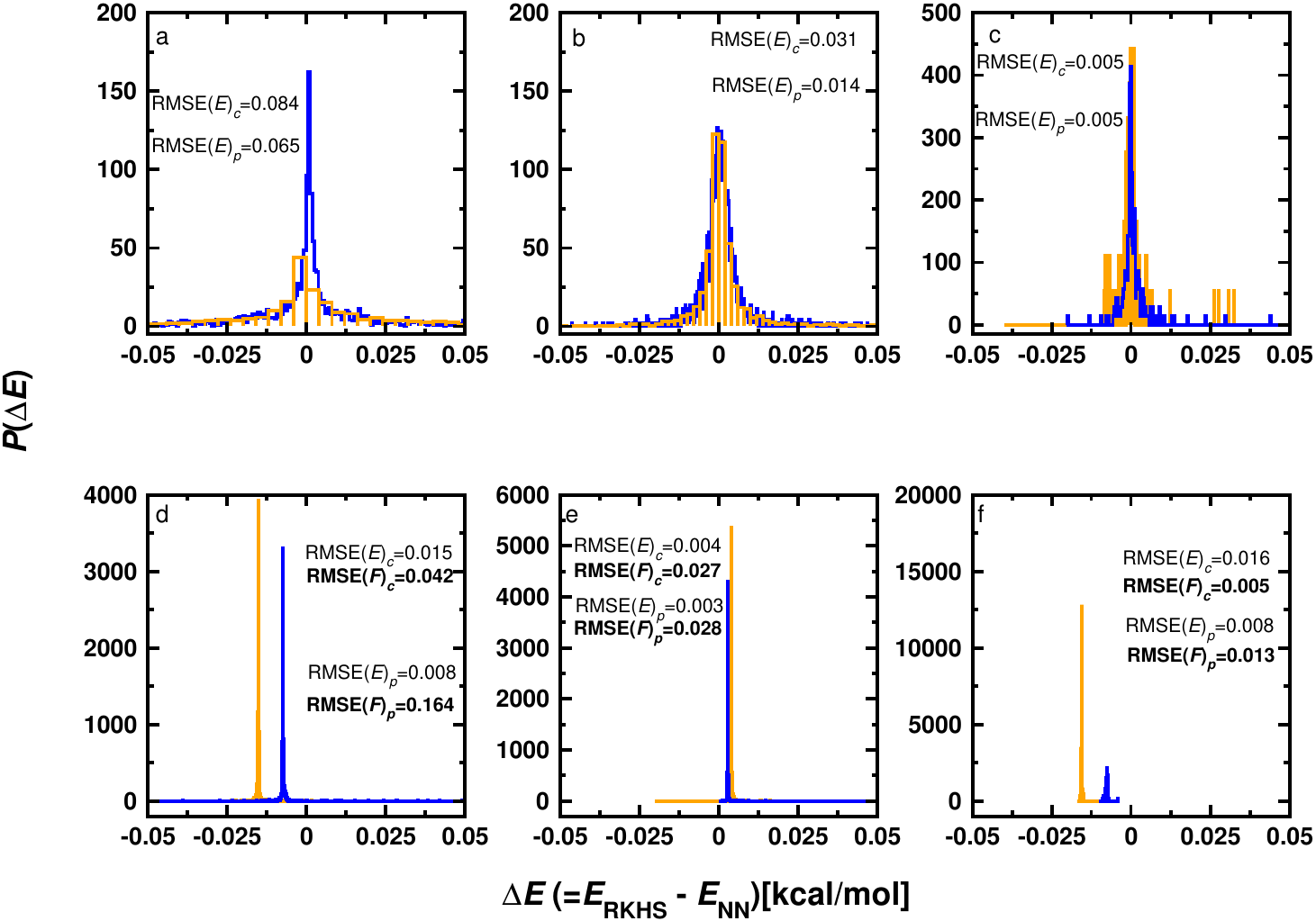}
    \caption{The probability distribution of the energy difference,
      $\Delta E = E_{\rm RKHS} - E_{\rm NN}$ for the test set for
      training set sizes of 500 (a, d), 1500 (b, e), and 3000 (c,
      f). Panels a to c: for ``energy-only'' training; panels d to f:
      for energy+force training. Clean (orange) energies and forces
      correspond to the RKHS energies and forces for H$_{2}$CO whereas
      the perturbed (blue) datasets refer to the sets obtained by
      adding random Gaussian noise with SD $=10^{-6}$ eV ($2.31\times
      10^{-5}$ kcal/mol) on energies and SD $=10^{-6}$ eV/\AA{}
      ($2.31\times 10^{-5}$ kcal/mol/\AA{}) on forces. For each
      dataset size, the data corresponds to the best of the two
      training calculations. RMSE($E$)$_{c}$ and RMSE($E$)$_{p}$
      represent the root-mean square error value in energy for the
      clean and 'noisy' datasets, respectively. RMSE($F$)$_{c}$ and
      RMSE($F$)$_{p}$ also have the same meaning but for the forces.}
    \label{sifig:fig2}
    \end{center}
\end{figure}

\begin{figure}
    \begin{center}
    \includegraphics*[scale=0.7]{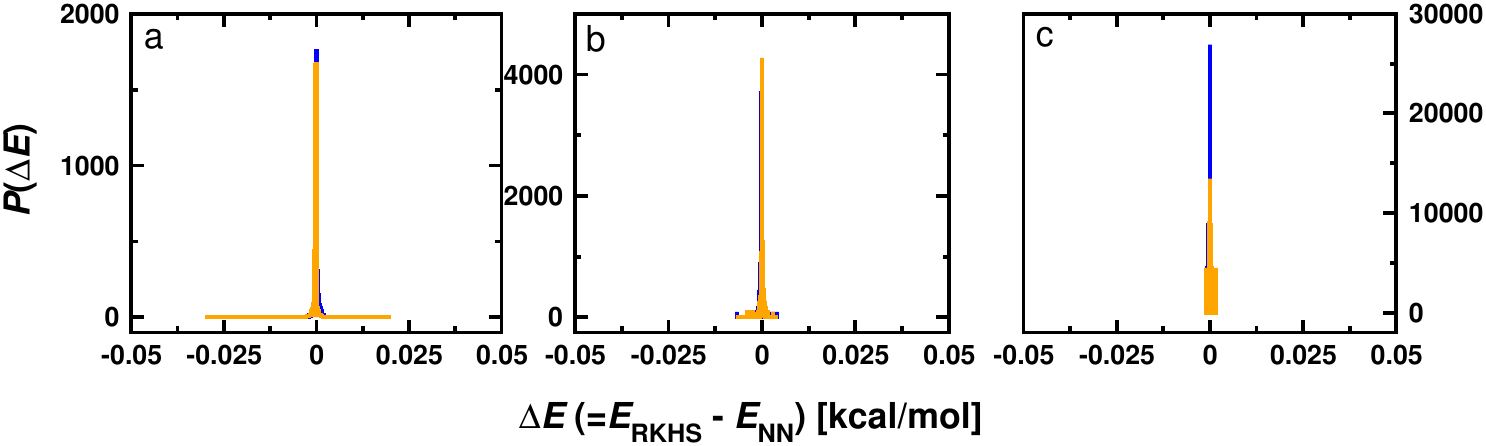}
    \caption{The probability distribution of the energy difference,
      $\Delta E = E_{\rm RKHS} - E_{\rm NN}$, for the H$_2$CO test set
      for training set sizes of $500$ (a), $1500$ (b) and $3000$
      (c). The trainings are for clean energy with clean force
      (orange) and clean energy with noisy (SD $=10^{-5}$ eV/\AA{})
      force data (blue) and with force weighting hyperparameter,
      $w_{\rm F}=1$. For training with $w_{\rm F} \sim 53$ see Figure
      \ref{fig:fig4}d to f.}
    \label{sifig:fig6}
    \end{center}
\end{figure}

\begin{figure}
    \begin{center}
    \includegraphics*[scale=0.7]{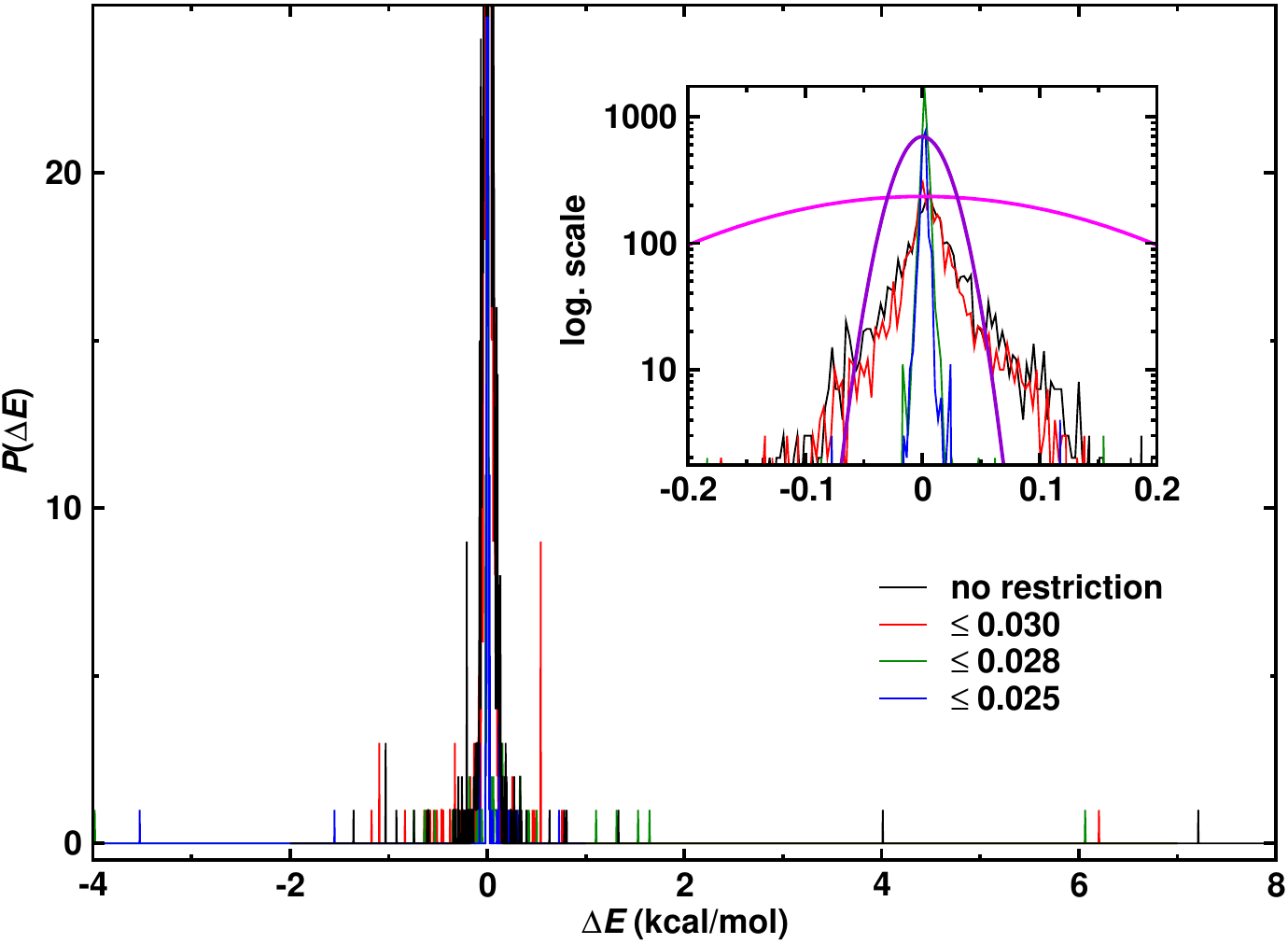}
    \caption{The probability distribution of prediction errors
      $P(\Delta E)$ for HONO trained on MP2 energies depending on the
      $T_1$ amplitude, see Figure \ref{fig:fig9}. The inset shows a
      close-up of the main figure and for the data without restriction
      (black) and $\leq 0.025$ (blue) on the $T_1$ amplitude Gaussians
      (magenta and violet, respectively) centered around $\Delta E =
      0$ show the aymmetric distribution of the error towards $\Delta
      E > 0$.}
    \label{sifig:fig5}
    \end{center}
\end{figure}

\begin{figure}
    \begin{center} 
      \includegraphics*[scale=0.6]{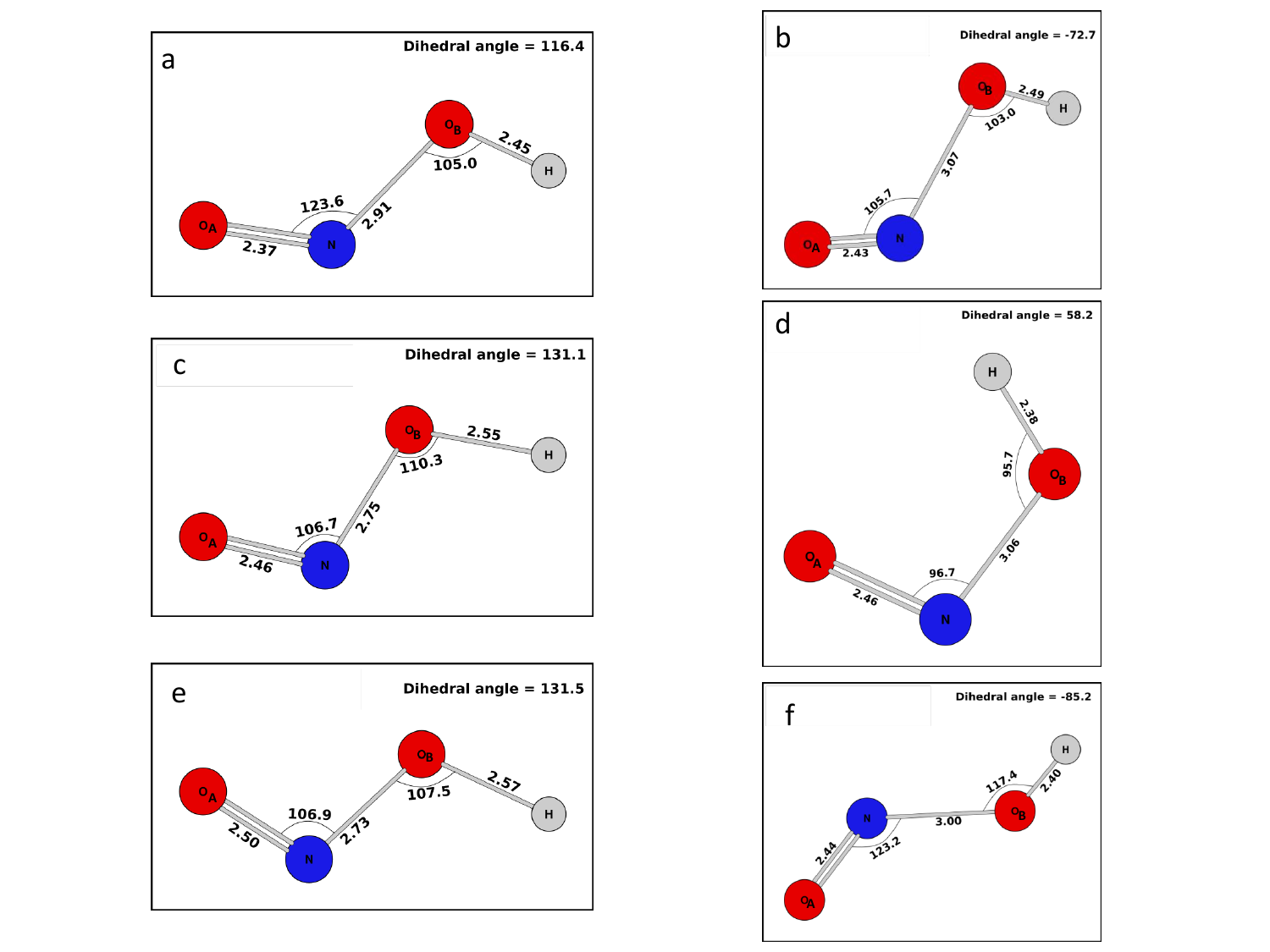}
    \caption{Structures for HONO encircled in Figure
      \ref{fig:fig11}b. The optimized(MP2/aug-cc-pVTZ)/equilibrium
      structure of \textit{trans-}HONO has $R_{\rm O_{B}H} =
      1.83a_{\rm 0}$, $R_{\rm O_{B}N} = 2.70a_{\rm 0}$, $R_{\rm
        NO_{A}} = 2.22a_{\rm 0}$, $\theta_{\rm HO_{B}N} = 101.6
      ^{\circ}$, and $\theta_{\rm O_{A}NO_{B}} = 110.7 ^{\circ}$ and
      $180.0^{\circ}$ dihedral angle. The six structures displayed are
      distorted significantly away from the minimum energy
      structure. Apart from the dihedral angle, the distortion takes
      place due to elongation of bonds and closing/opening up valence
      angles. For example, the structure in panel a has noticeably
      longer $R_{\rm O_{B}H}$ bond length and widely open $\theta_{\rm
        O_{A}NO_{B}}$. Furthermore, the structure in panel b has
      longer bond lengths for all three bonds, whereas the structure
      with $58.2^{\circ}$ dihedral angle has a narrow $\theta_{\rm
        O_{A}NO_{B}}$ angle.}
    \label{sifig:fig3}
    \end{center}
\end{figure}

\begin{figure}
    \begin{center} 
      \includegraphics*[scale=0.6] {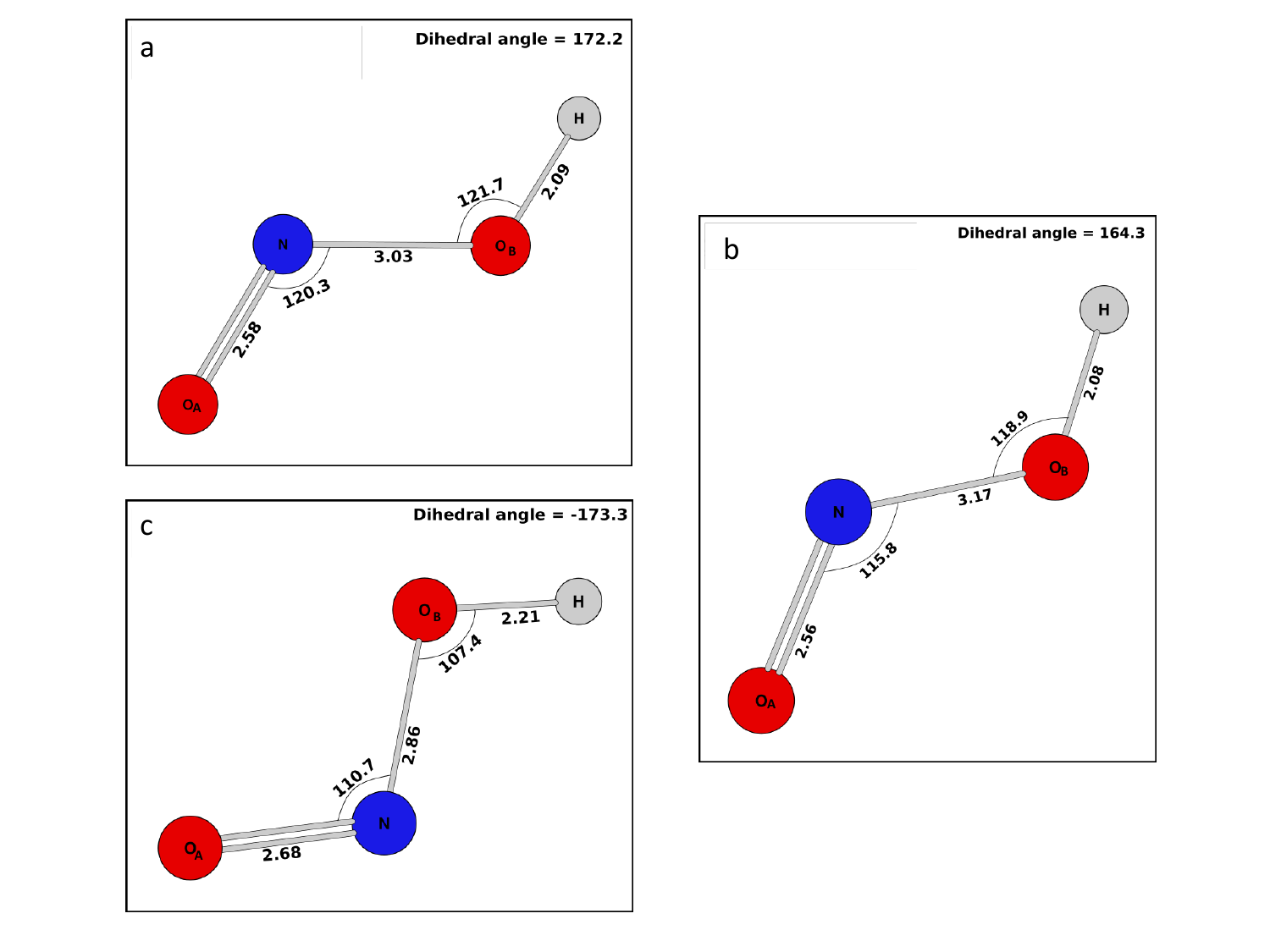}
    \caption{As Figure \ref{sifig:fig3} but for structures encircled
      in Figure \ref{fig:fig11}a. Unlike the structures shown in
      Figure \ref{sifig:fig3}, the dihedral angles are close to the
      optimized/equilibrium structure of \textit{trans-}HONO. The
      noticeable distortion occurs for these three structures due to
      longer bond lengths and opening up of bond angles. For example,
      both bond angles are wider for the structure shown in panel
      a. The structure of panel b has longer $R_{\rm NO_{A}}$ bond
      length and wider $\theta_{\rm HO_{B}N}$ bond angle.}
    \label{sifig:fig4}
    \end{center}
\end{figure}

\end{document}